%% file: main.tex
\documentclass[reprint,
twocolumn,
notitlepage,
superscriptaddress,
nofootinbib,
floatfix,
pra,
aps]{revtex4-2}

\input{./packages}
\input{./commands}
\input{./protocolenvironment}

\makeatletter
\let\cat@comma@active\@empty 
\makeatother

\begin{document}

\title{Anonymous and private parameter estimation in networks of quantum sensors}

\newcommand{\addressBerlin}{Electrical Engineering and Computer Science Department, Technische Universit\"at Berlin, Einsteinufer 17, 10587 Berlin, Germany}

\newcommand{\addressParis}{LIP6, CNRS, Sorbonne Universit\'e, 4 place Jussieu, F-75005 Paris, France}

\author{Jarn de Jong}
\affiliation{\addressBerlin}
\author{Santiago Scheiner}
\affiliation{\addressParis}
\author{Naomi R. Solomons}
\affiliation{\addressParis}
\author{Ziad Chaoui}
\affiliation{\addressBerlin}
\author{Damian Markham}
\affiliation{\addressParis}
\author{Anna Pappa}
\affiliation{\addressBerlin}

\begin{abstract}


Anonymity and privacy are two key properties of modern communication networks. In quantum networks, distributed quantum sensing has emerged as a powerful use case, with applications to clock synchronisation, detecting gravitational effects and more.
In this work, we develop a new protocol that, for the first time, combines the different cryptographic properties of anonymity and privacy for the task of distributed parameter estimation. That is, we present a protocol that allows a selected subset of network participants to anonymously collaborate in estimating the average of their private parameters. Crucially, this is achieved without disclosing either the individual parameter values or the identities of the participants, neither to each other nor to the broader network.

\end{abstract}

\maketitle

\section{Introduction}

The rapid advancement of quantum technologies promises a new era of secure communication, computing, data processing and sensing. In particular, quantum sensing allows for exceedingly precise and exact measurements, and has applications in civil engineering \cite{shettell2024geophysical} as well as for clock synchronisation \cite{komar2014quantum,komar2016,dai2020} and phase estimation~\cite{Knott2016,humphreys2013,Gagatsos2016}. Another cornerstone in modern quantum technologies is the development of quantum networks, i.e., infrastructures that leverage the unique properties of quantum mechanics to enable transmission of information with higher security or efficiency than their classical counterparts \cite{Chiribella:2009qjn,vanmeter12,KozlowskiWehner}. These quantum networks can be seen as part of the effort to design a future quantum Internet that promises unparalleled levels of robustness and security \cite{wehner2018}.

An interesting combination of the above leads to a network of spatially distributed agents, each holding a quantum sensor that measures some local parameter. In this setting, the agents aim to estimate a linear function over their parameters, commonly referred to as distributed parameter estimation. As in other networking tasks, malicious adversaries form an unavoidable threat that needs to be counteracted. In many applications (particularly in military, medical or even agricultural settings~\cite{edererTemperatureMonitoringAgricultural2023}) the network should be safeguarded against eavesdroppers who wish to learn the outcome of the estimation or the individual parameters themselves. This can be ensured by combining quantum sensing with quantum cryptographic methods \cite{shettell2022private,shettel2022metro,shettell2022cryptographic, Takeuchi2019,Okane2021, hassani2024privacy, Bugalho2025privaterobuststates,moore2025secure, huang2019cryptographic}. 

In~\cite{shettell2022private}, the authors present a protocol for private parameter estimation, which considers a quantum network of $\numberofagents{}$ agents, each hosting an individual parameter (e.g. as a property of a locally held sample). The protocol allows the agents to compute the desired linear function (such as the mean) of all parameters, without, importantly, any agent having to reveal their individual parameter to anyone else inside or outside the network. In other words, every member of the network can obtain information only about their own parameter and the global average. This notion of \emph{privacy} was further expanded and formalised in detail in \cite{hassani2024privacy,Bugalho2025privaterobuststates}.

Anonymity is another crucial aspect of modern communication.
Here anonymity refers to the goal that the \textit{identities} of communicating parties need to remain hidden rather than the information itself.
Anonymity was first considered in the quantum setting in \cite{Christandl2005}, where it was shown how to anonymously send quantum and classical messages, and has since been developed to more robust protocols \cite{brassard2007anonymous,unnikrishnan2019anonymity}, and guaranteeing anonymity in other scenarios such as conference key agreement \cite{hahn2020anonymous,grasselli2022secure}. Like covert sensing~\cite{tahmasbiSignalingCovertQuantum2021,haoDemonstrationEntanglementEnhancedCovert2022}, we obscure participation in the scheme. However instead of considering security against external adversaries we focus on security internal to a network, hiding information about other users’ behaviour. Our scheme is also device-agnostic as it is implemented on the communication level, as opposed to depending on the choice of physical infrastructure.

In this work, we combine the notions of privacy and anonymity to present the first protocol for \emph{anonymous private parameter estimation}. 
Specifically, we study the setting of a quantum network of $\numberofagents{}$ agents, who each receive one qubit of a \GHZtext{} state (from a potentially untrusted source). One special node, referred to as Alice, chooses the set of \emph{participants}, a subset of $\sizeofp \leq \numberofagents $ agents, from whose parameters she wishes to estimate the average. Apart from Alice, every agent in the network is aware of only their own role, but not which other agents are contained in the subset, or which agent is acting as co-ordinator (Alice). All nodes in the network together coordinate a scheme that allows Alice to learn the average, while maintaining the privacy requirement that individual parameters are kept secret, by exploiting the non-local correlations arising from the \GHZtext{} state.

This work builds on the idea of private parameter estimation from~\cite{shettell2022private} which we extend towards an anonymous setting using insights from ~\cite{hahn2020anonymous, grasselli2022secure, dejongAnonymousConferenceKey2023}. Care must be taken, however, when combining cryptographic properties; any part of the protocol involving parameter values or participant identities has the potential to compromise either privacy or anonymity respectively. Therefore, anonymous and private schemes cannot be straightforwardly combined in sequence, and one must ensure all desired properties are safeguarded at all times. Our efforts culminate in the first \APPE{} protocol and mark the first application of anonymity beyond the scope of secure communication within quantum technologies.
The significance of our contribution lies in adding anonymity to private parameter estimation while maintaining the same levels of accuracy and privacy. Further, we highlight that our protocol can easily be modified according to the required level of privacy or anonymity.

We structure our work as follows. We first give some background information on the task at hand, presenting the basic parameter estimation scheme in~\Cref{subsec:PPE}, and then explain the setting considered here in more detail in~\Cref{subsec: setting}. We then present our protocol in~\Cref{sec: protocol} detailing every step and sub-protocol. In~\Cref{sec: security}, we prove that our protocol is integrous (the parameter estimation can be trusted), private and anonymous. Finally, in~\Cref{sec:discussion}, we discuss adaptions to our security requirements as well as our protocol and conclude our work.

\section{Background}

In this section, we briefly describe the key mechanism for private parameter estimation in a quantum network, we introduce the anonymous setting we will consider in this work, and we describe the notions of privacy and anonymity that we will use.

\subsection{Private parameter estimation}\label{subsec:PPE}

Studies in quantum metrology have long suggested the use of large entangled states for the joint estimation of a linear function of locally held parameters~\cite{giovannetti_quantum-enhanced_2004, giovannetti_quantum_2006, giovannetti_advances_2011, eldredge2018optimal, komar2014quantum, degen_quantum_2017, proctor2018}, with early proof-of-concept experiments now being implemented~\cite{liu_distributed_2021, ho2024quantum}. The quantum advantage in this scheme comes from an asymptotically quadratic improvement in the number of probe-sample interactions (that is, the number of interactions that the local qubit, e.g. a single photon, must make with the material or field to be measured) \cite{Giovannetti2006, toth_quantum_2014, Gagatsos2016}.

Recent work has considered this task in a cryptographic setting, where these parameters should remain unknown to any other node of the network~\cite{shettell2022private, shettell2022private, hassani2024privacy, Bugalho2025privaterobuststates}. In other words, the different parties that form the network are able to collectively estimate a function of the local parameters, without the need to communicate this parameter to any other party. This principle is the basis of \emph{private parameter estimation}.

Analogous to~\cite{shettell2022private}, in this work, we focus on the estimation of a particular linear function of the local parameters: the mean. Such a functionality is in line with use cases such as clock synchronisation~\cite{komar2014quantum}, although it is possible to calculate other linear functions using different states as a resource~\cite{eldredge2018optimal, Bugalho2025privaterobuststates}.
In \cite{shettell2022private}, the process for the private estimation of the average value of the local parameters proceeds using a verified GHZ state across an $n$-user network. Each agent implements a local rotation $ \Lambda(\theta_i) = \ketbra{0}{0} + \mathrm{e}^{i\nicefrac{\theta_i}{n}}\ketbra{1}{1}$ to the received qubit, where $\theta_i$ is their private parameter, then measures it in the computational basis and announces the outcome.

The key idea of the protocol is that these individual announcements reveal no information about the private parameters, but allow the nodes to estimate $\bar{\theta} = \frac{1}{n}\sum_i \theta_i$ with high precision. More specifically the probability that the announcements have even parity in each round is $\frac{1}{2}\left( 1 + \cos(\bar{\theta}) \right)$. Repeated rounds therefore allow the network agents to gain a precise estimate of $\bar{\theta}$.

\subsection{Anonymous setting}\label{subsec: setting}

There are many reasons why anonymity may be required. Even in networks where every member is associated with a public identity, it may be necessary to obscure the relationships between users. For example, in medical use cases, a subgroup may be chosen according to particular characteristics, which participants may wish to remain private, or in political scenarios it may be preferred to keep secret who is involved in certain operations.

The goal of this work is to allow one agent, \textit{Alice} ($\alice$), to act anonymously as an orchestrator of a scheme where she can choose a set of $m \leq n$ participants, $\p$, and where she is  able to estimate $\avgp = (\nicefrac{1}{\sizeofp})\sum_{i\in\p} \theta_i$, the average of the local parameters of the agents in $\p$, while preserving both the anonymity of all agents involved and the privacy of the individual parameters.

Alice has full information regarding the identities of the network members, that is, she knows who is a participant and who is not. Regarding the parameters however, she is as constrained as any participant by the privacy conditions, meaning she will only have access to her own private parameter while estimating $\avgp$.

On the other hand, the property of \emph{anonymity} ensures that any other agent in the network, be it a participant in $\p$ or a non-participant in $\np$, will only know their personal role in the protocol (i.e. whether they are a participant or not) and the number of participants. They will however not know who Alice or the other (non-)participants are.  Regarding the parameters, they will only know their own private parameter. 

We utilise the definition of full anonymity from \cite{grasselli2022secure} and adapt it to our protocol and setting: we define anonymity in terms of closeness to an \emph{ideal} output state $\sigma$, which contains all of the quantum and classical information of the different parties. Ultimately, the definition ensures that the reduced state of $\sigma$ on any relevant subset $\g$ of agents in the network is independent of the choice of Alice and the participants.

\Cref{tab:information} summarises the information available to the agents of the network in this protocol.

\begin{table}[htbp]
    \centering
    \begin{tabular}{p{1.2cm}|p{1.2cm}p{1.8cm}p{1.2cm}p{0.5cm}p{0.5cm}p{0.6cm}p{0pt}}
        Agent $a_{i}$ & \centering  $a_{i} = \alice$ & \centering $\{a_{j} \in \p \}_{j \neq i}$ & \centering $|\p|=\sizeofp$ & \centering $\avgp$ & \centering $\theta_{i}$ & \centering $\theta_{j\neq i}$ & \\
        \hline
        $a_{i} = \alice$    & \yes  & \yes  & \yes  & \yes   & \yes & \no & \\
        $a_{i} \neq \alice$ & \no   & \no   & \yes  & \no    & \yes & \no & 
    \end{tabular}
    \caption{Information available to the different agents of the network. \textbf{Identities:} Alice ($\alice$), as the orchestrator, is the only one with information about who are the participants. All agents learn $\sizeofp$ (the size of $\p$). \textbf{Parameters:} all agents have access to their own parameter $\theta_{i}$, but none of them are able to get any information about other agents' parameters $\theta_{j\neq i}$. Only Alice can estimate the average of the participants' parameters, $\avgp$.}
    \label{tab:information}
\end{table}

\begin{figure*}[ht]
    \centering
    \includegraphics[width=1.0\textwidth]{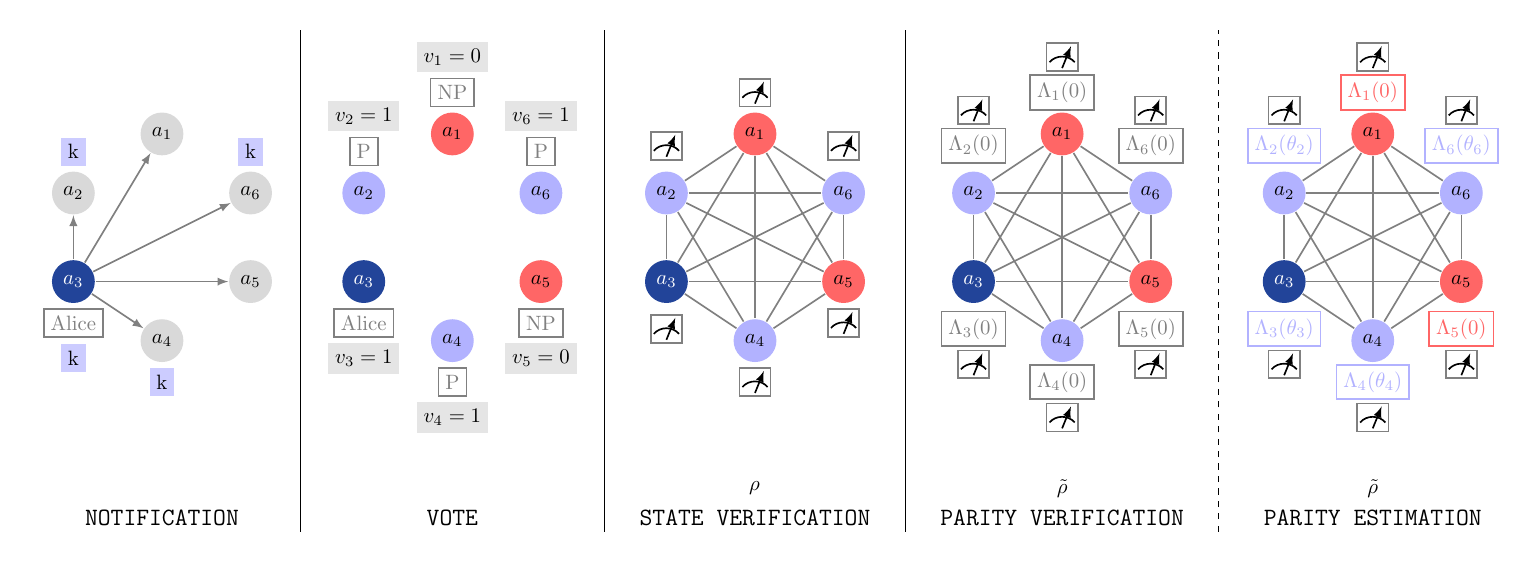}
    \caption{Illustration of~\APPE{}~on an example of a network of $\numberofagents=6$ agents, where $a_{3}$ is the co-ordinator Alice, $\alice$, and starts the protocol. In the first step, the nodes run \NOTIFICATION{} to allow Alice to anonymously notify the set of $\sizeofp=4$ participants $\{a_{2}, a_{3}, a_{4}, a_{6}\}$. In the second step the agents run \VOTE~for the participants to verify $\sizeofp$. During this round, participants vote anonymously $v_{2}=v_{3}=v_{4}=v_{6}=1$, and, in an honest scenario, non-participants vote $v_{1}=v_{5}=0$. 
    Next the network runs \STAVER{} to ensure that all agents share a state sufficiently close to a \GHZtext{} state. Using a secret key Alice can then communicate to the participants whether the state is used for \PARVER{} (\PARVERshort{}) or \PAREST{} (\PARESTshort{}). In \PARVERshort{}, everyone measures their qubit in the $X$-basis and it is used as trap round to verify that non-participants do not tamper with the desired estimation. In \PARESTshort{}, the participants first apply a rotation on their qubit using their private parameter and then measure in the $X$-basis, while non-participants are expected to just measure. By repeating \PARESTshort{} enough times, Alice can estimate $\avgp$.}
    \label{fig:steps}
\end{figure*}

\section{Protocol}\label{sec: protocol}

We now describe our protocol with all its sub-protocols in detail. 
The protocol starts by running the \NOTIFICATION{} sub-protocol, anonymously informing each participant whether or not they are included in $\p$. Crucially, each agent learns only their own status and gains no information regarding the inclusion of others.
This notification step can be implemented using classical protocols, such as those described in~\cite{broadbent2007information, hahn2020anonymous}; see also \Cref{appendix:sub-protocols}.

If only a single agent $a_i$ is included in $\p$, a malicious Alice can retrieve the private parameter of said agent. Specifically, when $\p=\{\alice, a_i\}$ the average value becomes $\avgp = \frac{1}{2} (\theta_{\alice}+\theta_{i})$. Since Alice knows $\theta_{\alice}$ and $\avgp$ (see \Cref{tab:information}) she can compute the private parameter $\theta_i$. The participants therefore need a guarantee that $\p$ is a large enough set. To this end, the network runs a self tallying majority voting protocol, \VOTE{}, e.g. from ~\cite{broadbent2007information}.
In the \VOTE{} protocol, agents in $\p$ input $1$ as their vote, and non-participants input $0$. The protocol counts the number of $1$ votes and reveals the number of agents in $\p$, $m$, to the network, allowing participants to abort if $m$ is not sufficiently large. Without loss of generality, we assume that non-participants act honestly in \VOTE{}, because the only scenario where cheating would be of advantage is if they collaborate with Alice, in which case it would be equally advantageous to simply include the dishonest agents in $\p$ (otherwise, Alice would notice that the total number of participants is incorrect, which constitutes a denial-of-service attack).

In order to be able to run the parameter estimation, the participants need to guarantee that the network shares a \GHZtext{} state. The agents therefore run \STAVER{} (\STAVERshort{}), a protocol that queries a source to distribute a \GHZtext{} which the network verifies. Suitable \STAVERshort{} protocols are presented in ~\cite{unnikrishnan2022verification, pappa2012}. 

\begin{figure*}[ht!]
    \centering
    \includegraphics[width=1.0\textwidth]{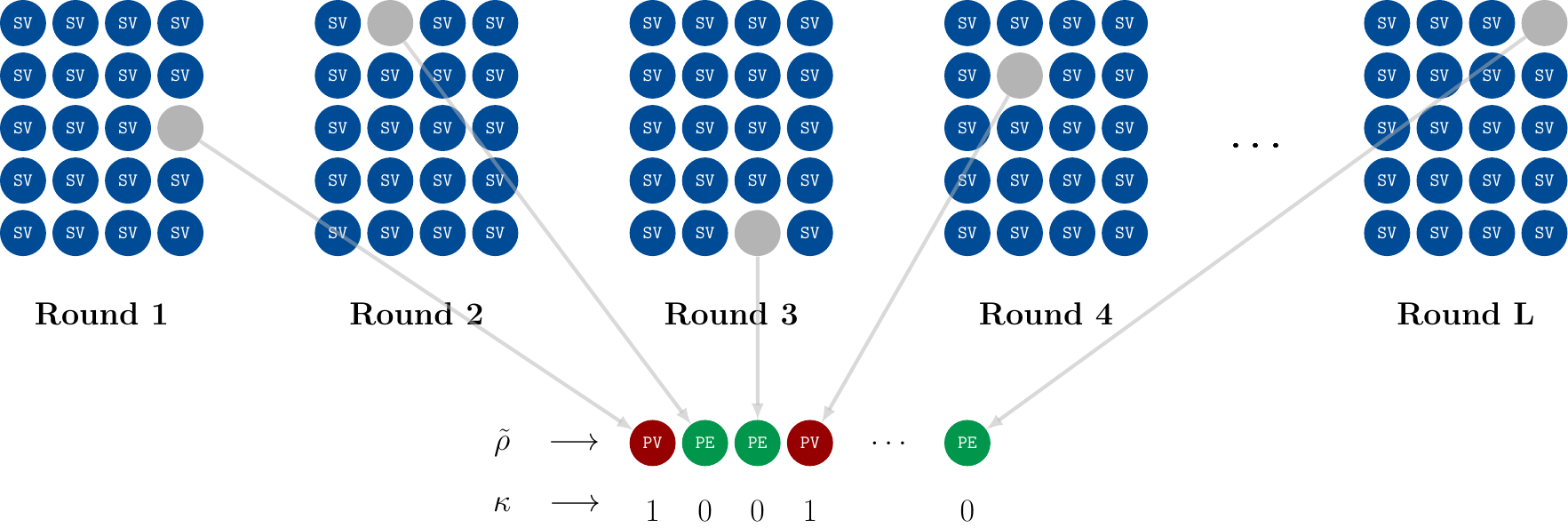}
    \caption{Illustrative example of the selection and use of target states $\tilde{\rho}$ from each round. In this example, five of the total $L$ rounds are shown. In each round a certain number of states are measured and tested to be \GHZtext{} states. Of the remaining states one is chosen to be the \emph{target state} $\tilde{\rho}$ highlighted in grey in the figure. The target state is then used for \PARESTshort{} or \PARVERshort{}, according to the key $\kappa$ generated during \ACKAshort{}.}
    \label{fig:rounds}
\end{figure*}

This is repeated $\numberoftotalrounds{}$    times, and each of these shared \GHZtext{} states is then (sequentially) used either for estimating \avgp{} (a `\PAREST{} (\PARESTshort{}) \emph{round}'), or to detect dishonest behaviour of non-participants (a `\PARVER{} (\PARVERshort{}) \emph{round}'). To coordinate these choices between the participants, they make use of a secret key $\kappa$: a bit string of length $\numberoftotalrounds$ with $k$ `1's at random positions indicating the \PARVERshort{} rounds, and $\nu = \numberoftotalrounds - k$ `0's indicating the \PARESTshort{} rounds, as illustrated in \Cref{fig:rounds}.
It is vital that the non-participants do not learn what rounds are \PARVERshort{} rounds, so in order to establish $\kappa$ the participants make use of a suitable \ACKA{} (\ACKAshort{}) protocol (see \cref{sec:discussion} for more details).

\begin{protocolfloat}[ht!]
\begin{protocol}{\PAREST{} (\PARESTshort{})}{Parameters $\{\theta_{i}\}_{i \in \{1,\dots,\numberofagents\}}$.}{Alice obtains parity estimation bit $\chi$.}
    \begin{protocollist}
        \item Each agent $a_i$ applies the unitary $\Lambda_{i}(\theta_{i}) = \ketbra{0}{0} + e^{i \frac{\theta_{i}}{\sizeofp}}\ketbra{1}{1}$ to their qubit.
        \item Each agent $a_{i}$ measures their qubit of the \GHZtext{} state in the $X$-basis and gets outcome $o_i$.
        \item Each agent except for Alice announces $o_i$. Alice announces a random bit.
        \item Alice computes and stores $\chi = \bigoplus_{i=1}^{n} o_i$.
    \end{protocollist}
    \label{prot:pe}
\end{protocol}
\end{protocolfloat}

In \PARESTshort{} (\Cref{prot:pe}) every agent $a_i$ applies the unitary $\Lambda_{i}(\theta_{i}) = \ketbra{0}{0} + e^{i \frac{\theta_{i}}{\sizeofp}}\ketbra{1}{1}$ to their qubit, with $\theta_i=0$ if $a_i\notin\p$. All agents measure their qubit in the $X$-basis and announce their outcome, except for Alice who announces a random bit. Over many runs, Alice can then use the parity of all measurement outcomes (including hers) to estimate $\avgp$, as the probability that the parity is even is $\frac{1}{2}\left(1 + \cos({\avgp})\right)$ (see~\Cref{subsec:PPE}).

\begin{protocolfloat}[ht!]
\begin{protocol}{\PARVER{} (\PARVERshort{})}{}{Alice obtains parity verification bit $\gamma$.}
    \begin{protocollist}
        \item Each agent $a_{i}$ measures their qubit of the \GHZtext{} state in the $X$-basis and gets outcome $o_i$.
        \item Each agent except for Alice announces $o_i$. Alice announces a random bit.
        \item Alice computes and stores $\gamma = \bigoplus_{i=1}^{n} o_i$.
    \end{protocollist}
    \label{prot:pv}
\end{protocol}
\end{protocolfloat}

With \PARVERshort{} (\Cref{prot:pv}) Alice implicitly verifies that non-participants act honestly. All agents measure their qubit in the $X$-basis and share their outcomes, except for Alice who again announces a random bit. Here, the total parity of all outcomes must always be $0$. Unintended behaviour of any non-participant, such as applying a unitary that would influence the estimation process, would disrupt this condition, allowing Alice to detect dishonest behaviour. Alice computes  $\delta$, the relative number of incorrect \PARVERshort{} rounds, which she can use to obtain a bound on the accuracy of the parameter estimation. 

Finally, Alice estimates $\avgp$ using the measurement outcomes of the \PARESTshort{} rounds and the fact that $\Pr\left(\chi_{j} = 0\right) = \nicefrac{\left(1 + \cos\left(\avgp\right)\right)}{2}$ for every $j$. She bounds its accuracy in terms of $\delta$, using~\Cref{eq: bias main text}.

The full protocol, which we denote by~\APPE{} (\APPEshort), is given in~\Cref{prot:appe} and a small example with 6 agents is illustrated in \Cref{fig:steps}.

As well as the aforementioned \GHZtext{} states, the protocol takes the secret identity of Alice and her selection $\p$ consisting of $m$ participants (potentially including herself) as an input.

\begin{protocolfloat}[h!]
\begin{protocol}
    {\label{prot:appe}\APPE{} (\APPEshort)}%
    {\GHZtext{} state source, parameters $\{\theta_{i}\}$, designated co-ordinator Alice and her choice of participants. Parameters $\nu$ and $k$.}%
    {Alice obtains accurate estimate of $\avgp$. Privacy and anonymity are maintained.}
    \begin{protocollist}
        \item Run \NOTIFICATION{} with \alice{} as coordinator.
        \item Run \VOTE.
        \item Each agent $a_i\notin\p$ sets $\theta_i=0$.
        \item Establish a biased secret key $\kappa$ of length $L$ between \p{}, with $k$ `$1$'s and $\numberoftotalrounds{} - \nu$ `0's. \label{protstep:makekappa}
        \item[] For round $1 \leqslant j \leqslant L$:
        \begin{enumerate}
            \item[5:] Distribute $\distributedghzstate$ $\GHZstate{}$ states.
            \item[6:] Establish a verified \GHZtext{} state through an \STAVERshort{}~protocol.\label{protstep:verifyGHZstates}
            \item[7-a:] If $\kappa_{j} = 0$: run \PARESTshort{}; Alice obtains and records outcome as $\gamma_{j}$.
            \item[7-b:] If $\kappa_{j} = 1$: run \PARVERshort{}; Alice obtains and records outcome as $\beta_{j}$.
        \end{enumerate}
        \item[8:] Alice computes the relative number of incorrect \PARVERshort{} rounds $\delta$ using $\{\gamma_{j}\}$.
        \item[9:] Alice estimates $\avgp$ using $\{\chi_{j}\}$, using the fact that $\Pr\left(\chi_{j} = 0\right) = \nicefrac{\left(1 + \cos\left(\avgp\right)\right)}{2}$ for every $j$.
    \end{protocollist}
\end{protocol}
\end{protocolfloat}

To ensure the integrity, privacy and anonymity of the protocol, there are certain expectations imposed on the end users, phrased as \emph{(resource) requirements}, which is common in cryptography. Some of these requirements arise from the \APPEshort{} protocol itself, and some are inherited from the sub-protocols. For both the main protocol and the sub-protocols, we assume that all classic channels are authenticated, a standard assumption. Furthermore, we assume that the network has access to an $\numberofagents$-partite, high fidelity \GHZtext{} state source.

The sub-protocols \NOTIFICATION{} and \VOTE{} rely on pairwise private channels, and \VOTE{} in particular relies on a simultaneous broadcasting channel. We note that no simultaneous broadcasting is necessary to announce the measurement outcomes in the main protocol, because Alice encrypts her outcome. Different implementations of \STAVERshort{} can rely on a public source of randomness, and the \ACKAshort{} protocol (\cref{protstep:makekappa} of \Cref{prot:appe}) will rely on assumptions as well (in particular, those discussed in \cref{sec:discussion} are either private pairwise classical communication, or sharing additional $\numberofagents$-partite, high fidelity \GHZtext{} states in the bounded storage model).

\section{Properties}\label{sec: security}

In this section, we define and discuss integrity, privacy, and anonymity. We provide proof outlines demonstrating how our \APPEshort{} protocol satisfies these properties and refer the reader to the appendix for complete and detailed proofs.

These security properties do not prevent denial-of-service attacks, which are possible at many stages. They also cannot prevent participants from inputting an incorrect value for their parameter (as this is still a legitimate use of the protocol). However, these security guarantees are not compromised (in the sense defined) under dishonest behaviour, or honest-but-curious behaviour, by any member of the network, eavesdropper, the source, or any collaboration of these.

\subsection{Integrity}

The \textit{integrity} of the protocol represents Alice's confidence that the output of the protocol (i.e., her estimate of $\avgp$, represented by the $\hat{\cdot}$ symbol) is both \emph{accurate} and \emph{precise}, even in the presence of malicious adversaries (represented by the $\cdot'$ symbol). These adversaries could include non-participants who should not be participating in parameter estimation, and yet wish to influence Alice's outcome. Accuracy is represented through constraining the bias, i.e. the difference between the expected and actual estimations produced by the protocol: $|\mathds{E}(\hat{\bar{\theta}}_{\p}') - \avgp|$ . The uncertainty is represented through any effective change to the variance, $|\Delta^2 \hat{\bar{\theta}}_{\p}' - \Delta^2 \hat{\bar{\theta}}_{\p}|$.

Central to this guarantee is the use of \PARVERshort{}, which is enabled by the secret key shared by the participants, and hence the security of \ACKAshort{} is also a vital consideration. We use these verification rounds to ensure that the only individuals able to influence \PARESTshort{} rounds are the participants, i.e., an adversary cannot impact a round of \PARESTshort{} without this being recorded by a wrong outcome in \PARVERshort{}. In particular, note that the outcome of \PARESTshort{} is a bit string, which is then used to estimate $\avgp$; we show that there is an exponentially small probability of adversaries remaining undetected in \PARVERshort{} rounds, relative to the number of \PARESTshort{} bits that they are able to flip. 

Given that we can limit any perturbation that an adversary can enact on the bitstring output by \PARESTshort{}, we then show that we can similarly constrain the bias of the estimation of $\avgp$. More specifically, when $\nu$ rounds are used for \PARESTshort{} and $k$ rounds are used for \PARVERshort{}, we limit the bias by:
\begin{equation}\label{eq: bias main text}
\begin{split}
    \Pr(|\mathds{E}(\hat{\bar{\theta}}_{\p}') - \avgp| \geq \eta ) \\ \leq \exp(-2 (f(\eta, \theta)- \delta)^2 \frac{\nu k^2}{(k + \nu)(k + 1)})
    \end{split}
\end{equation}
where $f(\eta, \theta)$ is a polynomial in $\eta$ and $\theta$, and $\delta$ is the proportion of \PARVERshort{}~rounds which give the outcome 1. This shows that there is an exponentially low probability of an attack causing a bias without causing a similarly significant response in the~\PARVERshort{}~rounds.

Furthermore, we show that this attack in fact has no effect on the expected variance of the estimation of $\avgp$, that is:
\begin{equation}
    |\Delta^2 \hat{\bar{\theta}}_{\p}' - \Delta^2 \hat{\bar{\theta}}_{\p}| = 0.
\end{equation}
Further details are given in \Cref{appendix:integrity}.

\subsection{Privacy}

\emph{Privacy} is understood as the ability of any agent to contribute to the estimation of a global linear function of local parameters $\{\theta_i\}$, such that any subset of dishonest agents $\dis$ in the network obtains no more information about any parameter $\theta_i$ than the information they already have from knowing the global parameter, the local parameters of the agents in the dishonest set $D$, and any function of these values. The notion of privacy in this work is relevant only during \PARESTshort{}, as this is the only stage in which the parameters are used to estimate the average of the participants' parameters.

The proof for this property follows from~\cite{shettell2022private, hassani2024privacy,Bugalho2025privaterobuststates}. The core idea is that a
set of agents can \emph{privately} estimate a global function of local parameters if the quantum Fisher information matrix that depends on the state, after parameter encoding, is a rank-1 matrix. This means that the target linear function can be estimated with arbitrary precision, while no information about any other function of the local parameters can be extracted from the system.

In particular, this work is concerned with the mean of the local parameters. This function can be represented by $\mathbf{w}=(1/m, \dots, 1/m)^{T}$, a vector of weights \cite{proctor2017networked}, so that:
\begin{align}
    \avgp = \mathbf{w}^{T} \vecp =
    \left( \frac{1}{m}, \dots, \frac{1}{m} \right)
    \begin{pmatrix}
    \theta_{1} \\
    \vdots \\
    \theta_{m}
    \end{pmatrix},
\end{align}
which means that $\mathbf{w} \mathbf{w}^{T}$ is a constant matrix.
This, together with the continuity relation of the Fisher information, implies that any state $\rho_{\vecp}$, the state after the local encoding of the parameters, that satisfies
\begin{align}
    \label{eq:condition-average-main}
    \partial_{i} \rho_{\vecp} = \partial_{j} \rho_{\vecp}, \qquad \forall i,j
\end{align}
can be used for the private estimation of linear functions of the local parameters.

Furthermore, for the case of the mean of the local parameters $\avgp$ it is shown that for an $\numberofagents$-partite \GHZtext{} state and encoding unitaries $\Lambda(\vecp) = \bigotimes _{i=1}^{\numberofagents}(\ket{0}\bra{0} + e^{i \nicefrac{\theta_{i}}{\sizeofp}}\ket{1}\bra{1})$, $\avgp$ is the only information that can be retrieved from the resulting state:
\begin{equation}
    \GHZstatetheta=\frac{1}{\sqrt{2}}\left(\ket{0}^{\otimes \numberofagents}+ e ^{i \avgp}\ket{1}^{\otimes \numberofagents}\right).
    \label{eq:GHZ-Ave-n}
\end{equation}

More generally, for an ideal state $\sigma$, a state verification scheme produces an output $\rho$ with the guarantee that $\Pr\left(\|\sigma - \rho\|_{\mathrm{tr}} \geqslant \varepsilon_{\STAVERshort{}}\right) \leqslant \alpha$, for some security parameter $\varepsilon_{\STAVERshort{}}$ and some confidence bound $\alpha$ ($\|\cdot\|_\mathrm{tr}$ denotes the trace norm). In \cite{hassani2024privacy}, `$\varepsilon$-privacy' is used to quantify privacy, introducing $\varepsilon_{\mathrm{priv}}$ (defined in terms of the coefficients of the Quantum Fisher Information matrix) which is associated with the maximum amount of information that can be extracted about each local parameter $\theta$. From this definition, and using the outcome of state verification, we can bound the leakage of information as:
\begin{align}
    \varepsilon_{\text{priv}} \leq 2 \, \varepsilon_{\STAVERshort{}}
\end{align}
with probability $1-\alpha$.

The detailed proof schemed in this subsection can be found in \Cref{appendix:privacy}.

\subsection{Anonymity}\label{sec:prop-anon}
We prove the anonymity of our protocol under an adapted version of \emph{full anonymity} as defined in \cite{grasselli2022secure}, which requires that an ideal output state of the \APPEshort{} protocol over all relevant registers (containing both classical and quantum information) needs to be \emph{fully ideally anonymous} - a definition that captures the notion that the state is independent of the choice of sender and participants, for every subset of network agents $\g \subset \{ a_{i}\}_{i=1}^{n}$ that does not include the sender. Anonymity is then quantified by the $\varepsilon_{a}$-closeness of the actual output state to such an ideal output state, which we call $\ro$ and $\sout$, respectively. 

Our protocol consists of several sub-protocols that all contribute differently to the final output state of the \APPEshort{} protocol and play different roles for anonymity. We can implicitly model the relevant steps of our protocol as a CPTP map $\prch$, so that $\ro = \prch \left(\ri\right)$ for some input state $\ri$, and $\sout = \prch \left(\si\right)$, where $\si$ is the ideal input state defined as:
\begin{equation}
    \si= \GHZstatethetaketbra^{\otimes L}_{\quantreg} \otimes \sigma_{C}\otimes \sigma_E,
\end{equation}
with $R$ the register that contains the quantum states shared in the network after \cref{protstep:verifyGHZstates}, $C$ a classical register containing a transcript of all public communication and $E$ the quantum register of Eve.

To prove the anonymity of our protocol we first show that this ideal output state $\sout$ satisfies our definition of a fully ideally anonymous state (\Cref{def:anonymity} in the appendix). Subsequently, the $\varepsilon_{a}$-anonymity defined in \Cref{def:epsanonymity} in the appendix then follows from the fact that the states after \cref{protstep:verifyGHZstates} are verified \GHZtext{} states. 

Indeed, any state verification scheme ensures that
$ \Pr\left(\|\ri - \si\|_{\mathrm{tr}} \geqslant \varepsilon_{\STAVERshort{}}\right) \leqslant \alpha,
$
for some security parameter $\varepsilon_{\STAVERshort{}}$ and some confidence bound $\alpha$ that decreases with growing $\distributedghzstate$.

By the data processing inequality, we also have $\|\prch(\ri)-\prch(\si)
\|_\mathrm{tr}\leq\|\ri-\si\|_\mathrm{tr}$ which immediately implies that  
\begin{equation}
    \|\ro-\sout\|_{\mathrm{tr}}\leqslant\varepsilon_a
\end{equation}
holds for $\varepsilon_a=\varepsilon_{\STAVERshort{}}$ with probability $1-\alpha$. Our protocol is therefore $\varepsilon_a$-anonymous by \Cref{def:epsanonymity}.

We refer to \Cref{appendix:anonymity} for more details, but note that we have omitted the confidence window there, essentially taking $\alpha = 0$. 

\begin{remark}
    While privacy and anonymity are two cryptographic properties of a protocol, neither is a functionality in itself. There are thus no protocols that simply achieve anonymity or privacy but rather we design protocols for a certain task - in our case parameter estimation - that safeguard these properties. In this work these properties coexist independently of one another, while both directly depend on the accuracy of the \GHZtext{} state verification protocol.
\end{remark}

\section{Discussion}\label{sec:discussion}
In this work we contribute to the growing number of cryptographic protocols for quantum sensing applications. We specifically manage to improve the private parameter estimation protocol presented in~\cite{shettell2022cryptographic} in order to also provide anonymity, while at the same time maintaining integrity and privacy. Indeed, we have presented the first protocol that allows an agent in a network to estimate the average of the parameters of some chosen subset of agents, while their identities as well as the parameter values, remain hidden. 

Nevertheless, several challenges remain. Most notably, the protocol heavily relies on verified \GHZtext{} states obtained through state verification, which implies the utilisation of a quantum memory to prevent loss of anonymity, or leaking private information. Moreover, many methods for creating and distributing \GHZtext{} states suffer both in success rate and robustness to noise \cite{webbExperimentalAnonymousQuantum2024,thalackerAnonymousSecretCommunication2021} when the number of agents increases; our anonymous setting requires \GHZtext{} states shared between the entire network rather than only sharing them between the participants as in \cite{shettell2022private}.
The possible anonymity and privacy guarantees that can be achieved directly depend on the quality of \GHZtext{} states that can be assumed through state verification.
These considerations pose practical limitations that must be addressed with improvements in the actual protocol and in future implementations.

Towards such improvements, consider the fact that the protocol and its proofs (as presented in \cref{appendix:integrity,appendix:privacy,appendix:anonymity}) rely on the fact that the \GHZtext{} state is verified; the guarantees on both privacy and anonymity are derived from this. However, improved proof techniques could alleviate this strong requirement, instead ensuring the privacy and anonymity from (the announcements during) \PARVER{} instead\footnote{Indeed, compare with modern entanglement-based QKD protocols. There, instead of obtaining security by verifying that Alice and Bob share an EPR pair $\ket{00} + \ket{11}$, they merely cross-compare some of their measurement results to verify the security of their key through e.g.~entropic uncertainty relations, thereby improving their key rates with orders of magnitude.}.
Note that such proof techniques could drastically improve the efficiency of our protocol in terms of both the number of necessary \GHZtext{} states that need to be distributed, and the size of the necessary quantum memory. Additionally, \VOTE{} relying on simultaneous broadcasting is another challenge, because this may not always be feasible in realistic network settings.
Future work should explore alternative approaches to mitigate these constraints, ensuring that the protocol remains both scalable and practical.

There are also various methods for the participants to establish the secret key $\kappa$ during \cref{protstep:makekappa} of \cref{prot:appe}. Essentially, to establish $\kappa$ the participants need to run a fully anonymous \ACKA{} (\ACKAshort{}) scheme, e.g.~in \cite{hahn2020anonymous,grasselli2022secure}. Note that the participants do not need to run \ACKAshort{} to create a shared secret key of length \numberoftotalrounds{}: because $\kappa$ contains only a fraction $\nicefrac{k}{\numberoftotalrounds{}}$ of `1's, it can be compressed to a bit string of length $h_{2}\left(\nicefrac{k}{\numberoftotalrounds{}}\right) \cdot \numberoftotalrounds{}$ (where $h_{2}\left(\cdot\right)$ denotes the binary entropy). The participants only need to establish a secret key of that length, after which they can individually `decompress' it to the desired length \numberoftotalrounds{}.

As presented, the protocol allows \alice{} to estimate the average of the parameters $\{\theta_{i}\}_{i \in \p{}}$. By having all agents change their local rotation $\Lambda(\theta_i)$ to $\Lambda(\theta_{i}, a_{i}) = \ketbra{0}{0} + \mathrm{e}^{i\nicefrac{a_{i}\theta_i}{n}}\ketbra{1}{1}$ (with $a_{i} \in \mathbb{R}$), \alice{} can estimate any linear function $\sum_{i \in \p} a_{i} \theta_{i}$ instead.
However, this approach only works when the agents are aware of their weights $a_{i}$, and when the agents are able to change their local rotation. In contrast, a fixed interaction might prove more relevant in a quantum sensing setting, so that the local rotation cannot be adapted. This was addressed in \cite{Bugalho2025privaterobuststates}, which departed from utilising \GHZtext{} states to other multi-partite entangled states: the state that is distributed is carefully adapted by the source to reflect the weights $a_{i}$, while maintaining the privacy of all the agents. Future research has to determine whether these states can also be used in our scheme, additionally safeguarding the anonymity of the involved parties.

Beyond adaptations for generic linear functions, adjustments to the protocol can be made in terms of the anonymity. Our definition of anonymity, as adapted from \cite{grasselli2022secure}, is called \emph{full}, and can be understood as the most stringent form of anonymity where (except for \alice{}) no agent is aware of anyone else's role; \NOTIFICATION{} and \VOTE{} are necessary in our protocol because of this requirement. 
If the identity of \alice{} is known within the network and she is trusted, \VOTE{} can be omitted because the participants do not need an extra guarantee for the privacy of their parameters.
Moreover, in such a setting \NOTIFICATION{} can be replaced by simple private pairwise communication between \alice{} and all other agents. 
Alleviating the anonymity even further, one arrives at \emph{partial} anonymity, where the set of participants is completely aware of each other. Note that in this setting it is considerably easier to obtain the shared secret key because \emph{partially anonymous} protocols are easier to implement \cite{grasselli2022secure,dejongAnonymousConferenceKey2023}; it is even customary to assume that a pre-shared key is already in place. 

Given the commercial and societal promise of quantum sensors, as well as the growing interest in utilising large scale quantum networks, function estimation is particularly appealing for combining the cryptographic and metrological advantages of quantum correlations, with recent proof-of-concept experimental demonstrations of some of the protocols considered in this work~\cite{thalackerAnonymousSecretCommunication2021, ho2024quantum,webbExperimentalAnonymousQuantum2024} or their adaptations \cite{Rueckle2023ExperimentalAnonymous}. Despite the remaining challenges in its implementation, the protocol described in this paper extends the functionality of the scheme, thereby increasing its practicality and security and bringing us closer to profiting from quantum networks at scale.

\section*{Acknowledgments}
We thank Sean W. Moore for useful discussions.

JdJ, SS, NRS, DM \& AP acknowledge the Quantum Internet Alliance (QIA), which has received funding from the European Union's Horizon 2020 research and innovation programme under grant agreement No 820445 and from the Horizon Europe grant agreements 101080128 and 101102140.
SS, NRS \& DM acknowledge the PEPR integrated project EPiQ ANR-22-PETQ-0007 part of Plan France 2030. NRS \& DM acknowledge support from the ANR project QNS ANR-24-CE97-0005-01.  AP acknowledges support from the Emmy Noether DFG grant No. 41829458. ZC and AP acknowledge funding from the Hector Fellow Academy.

\bibliography{bibliography}
\vfill \pagebreak \onecolumngrid

\appendix

\section{Notation}
\label{appendix:notation}

\newcommand{\itemdef}[2]{\item[$#1$] \hspace{1.5em} \parbox[t]{0.85\linewidth}{#2}}

\noindent
\begin{itemize}
    \itemdef{\alice}{Alice/co-ordinator}
    \itemdef{\p}{Set of participants}
    \itemdef{\np}{Set of non participants}
    \itemdef{\numberofagents}{Size of network, $n = |\p \cup \np|$}
    \itemdef{\sizeofp}{Size of set of participants, $|\p|$}
    \itemdef{a_i}{$i$-th agent (node) in the network}
    \itemdef{\theta_i}{Parameter of $a_i$}
    \itemdef{\vecp}{Vector of all parameters, $\vecp = (\theta_1, \dots, \theta_n)$}
    \itemdef{\vecp_\p}{Vector of participants' parameters}
        \itemdef{\avgp}{Average of participants' parameters, $\bar\theta_\p~=~\frac{1}{\sizeofp} \sum_{\{i : a_i \in \p\}} \theta_i$}
    \itemdef{\hat{\bar{\theta}}_{\p}}{Estimation of $\avgp$; for readability we also use notation $\hat{\theta} := \hat{\bar{\theta}}_{\p}$}
    \itemdef{\hat{\bar{\theta}}'_{\p}}{Estimation of $\avgp$, after it has been perturbed by adversarial behaviour; we also use notation $\hat{\theta}' := \hat{\bar{\theta}}'_{\p}$}
    \itemdef{L}{The total number of states used in \APPEshort{}~(corresponding to the number of states to be produced by \STAVERshort{})}
    \itemdef{\nu}{The number of rounds of \PARESTshort{}~which are carried out (expected to be $L - k$, unless some states are discarded)}
    \itemdef{k}{The number of rounds of $L$ used for \PARVERshort{}}
    \itemdef{\delta}{The proportion of incorrect \PARVERshort{}~rounds}

\end{itemize}

\section{Integrity}
\label{appendix:integrity}

Integrity refers to the ability to accurately calculate the objective function -- that is, to retain the desired functionality of the protocol -- even in the presence of malicious adversaries. This is closely related to the \textit{soundness} of the protocol, i.e. the ability to detect any malicious activity. In general, we follow the notation and definitions from~\cite{shettell2022cryptographic}.

More precisely, the actual mean to be calculated is $\overline{\theta}_{\p}$, and the correct functioning of the protocol produces an estimate to this, $\hat{\overline{\theta}}_{\p}$ (for readability, for the remainder of this section, we will instead use the notation $\hat{\theta}$ to represent the estimate to the mean). The protocol is unbiased if the expected estimation of the mean matches the true mean, $\mathds{E}(\hat{\theta}) = \overline{\theta}_{\p}$. In general, a realistic implementation of the protocol with some adversaries will produce a new estimate $\hat{\overline{\theta}}'_{\p}$ (similarly, we will use the notation $\hat{\theta}'$ for the remainder of this section). We are hence interested in the bias:
\begin{equation} \label{eq: bias def}
    |\mathds{E}(\hat{\theta}') - \overline{\theta}_{\p}|.
\end{equation}
We are also concerned with a measure of the uncertainty introduced by malicious parties~\cite{shettell2022cryptographic}:
\begin{equation}
    |\Delta^2 \hat{\theta}' - \Delta^2 \hat{\theta}|.
\end{equation}

By the expected behaviour of the protocol, members of $\p$ have complete freedom to input any angle from 0 to $2 \pi$ at each parameter estimation round (although the desired behaviour is to enter a value in the range 0 to $2\pi/m$, this is not technically enforced), and hence in this analysis we do not consider the dishonest behaviour of participants, who could freely produce any bias or uncertainty. Instead, we consider dishonest behaviour from any subset of the non-participants, or another adversary, given that any of these may have control over the source, or quantum channels between the source and members of the network.

\subsection{Constraining the affected rounds of \PAREST} \label{sec: parest bitflips}

The estimate $\hat{\theta}$ is produced within the final stage of \APPEshort{}. During this step, a bitstring is created of length $L$, of which we expect that $k$ are used for \PARVERshort{}, which produces error rate $\delta$. The remaining $\nu = L-k$ bits are used for \PARESTshort{}. The subset of rounds, $V$, which are used for \PARVERshort{} are decided by a preshared key between the participants. We use an \ACKAshort{} protocol that has an exponentially low probability of failure, and therefore we will at first assume that the adversary has no knowledge of which bits of this block are are used for \PARESTshort{}~and which are used for \PARVERshort{}. Therefore, any attack is permutation invariant with regards to the $L$ bits of \APPEshort{} -- it has equal chance of landing on a \PARESTshort{}~or \PARVERshort{}~round. 

We also assume that we start the \APPEshort{} protocol with a \GHZtext{} state distributed across the whole network. This is based on the correct functioning of \STAVERshort{}, with the assumption that the source of randomness to decide which rounds are used for verification is called after the state distribution.

The outcome of all rounds of \PARVERshort{}~can be expressed as a test function, $\{0,1\}^{k} \rightarrow \{ \checkmark, \varnothing \}$, where the outcome is $\varnothing$ if $\sum_{i=1}^k v_i > \delta k$ and $\checkmark$ otherwise, where $v_i$ is the parity of an individual verification round (computed by Alice) and $\delta$ represents some accepted level of error (that is, the proportion of verification rounds that fail, where the overall test still passes). If more than the proportion $\delta$ of these rounds give parity 1 ($v_i = 1$), the protocol is abandoned. Alternatively, Alice can place no requirements on $\delta$, but take this as an outcome of the \PARVERshort{} rounds and use it in calculating the expected bias of her estimate.

Consider the situation where, unknown to the adversaries, all of the participants input $\theta_i = 0$, but otherwise behave honestly (that is, they simply make $X$ measurements and announce the outcome). We can use the following result from~\cite{tomamichel2017largely} (proof omitted):
\begin{lemma} \label{lemma: verification failure}
    Consider a set of binary random variables $Z = (Z_1, Z_2, ..., Z_L)$ where $L = \nu + k$. Let $V$ be an independent, uniformly distributed random subset of size $k$. Then:
    \begin{equation} \label{eq: verification failure}
        \Pr(\sum_{i \in V} Z_i \leq k \delta \wedge \sum_{i \in \bar{V}} Z_i \geq (L-k) (\delta + \omega))  \leq \exp(-2 \omega^2 \frac{(L-k) k^2}{L(k + 1)}).
    \end{equation}
\end{lemma}
This is applied such that the random variables $Z$ are the outcomes of each round of \APPEshort{}, where the subset $V$ are used for verification and $\bar{V}$ are used for parameter estimation. The correct outcome for each round in both verification and estimation is 0, but any bit may be flipped to a 1 by the action of adversaries. However, as the adversaries do not know which rounds are in $V$, then each $Z_i$ has the same distribution.

The term $\sum_{i \in V} Z_i \leq k \delta$ then specifies the case where the protocol passes. It is to be expected that up to $\delta \nu$ of the parameter estimation rounds come out as 1 (this is the accepted error). However, we are interested in a further bias that affects a further $\alpha \nu$ of the remaining outcomes. The probability of this occurring when the protocol passes is exponentially small (as shown in Lemma~\ref{lemma: verification failure}).

This result is still useful in the case that $\theta_i \neq 0$. Note that, using Lemma~\ref{lemma: verification failure} we can limit (by $\alpha$) the proportion of rounds in which an adversary behaves in such a way that, should the compromised quantum state or classical information be used for verification, the output bit would be 1. Thus, we need to confirm that any state which would pass a verification round can be used effectively for parameter estimation (that is, giving 0 or 1 with the correct probability). 

Now we consider the sorts of attacks that can be carried out by dishonest participants. Regarding classical communication, the honest participants publicly announce their measurement result, and Alice keeps her measurement result private and obscures it by announcing a random bit, and hence the only attack that can be carried out using only the classical communication is to flip the outcome bit. Alternatively, simultaneous broadcast could be used, in which case the adversaries again have no knowledge of the parity of the rest of the bit string before announcing their own bit.

Now we consider any attacks on the quantum state, starting with the assumption of a distributed \GHZtext{} state. As the local operations performed by the parties commute, we can assume that Alice and the participants, as well as honest non-participants, have all made their $X$ measurements and now have the appropriate outcome. The state across the $l$ remaining members of the network is therefore:
\begin{equation}
     \frac{1}{\sqrt{2}}\left(\ket{0}^{\otimes l} + (-1)^{h} e^{i \avgp} \ket{1}^{\otimes l}\right)
\end{equation}
where $h$ is the parity of all of the $X$ measurement outcomes of honest members of the network, $\mathcal{H}$ -- which is unknown to the dishonest users, as $\mathcal{H}$ contains Alice, who obscures her outcome. We must assume that the dishonest users have knowledge of $\avgp$, as many rounds may have already occurred, or the protocol may have been run previously.

Given that $h=0$ or $1$ with equal probability, we can see that the density matrix of this state is independent of $\avgp$ (indeed it is $\frac{1}{2}\mathds{1}$, the maximally mixed state, in the basis spanned by $\ket{0}^{\otimes l}, \ket{1}^{\otimes l}$). Therefore any rotations or measurements made locally among the adversaries has the same impact on the classical information (and in particular the only relevant information, the parity of their shared $l$-bit string) for any value of $\avgp$ -- that is, any behaviour that causes a bit flip of the parity in a round of parameter estimation does the same for parameter verification.

This situation is different if Alice does not encode her outcome. In this case, the dishonest non-participants could potentially force a particular outcome (e.g. parity 0 for a round), which would not be detected by the verification scheme, which would damage integrity (or at least present a denial-of-service attack). Hence, in this case simultaneous broadcast should be enforced.

\subsection{Bias}

We would like to use Lemma~\ref{lemma: verification failure} to bound the potential bias of $\hat{\theta}'$ (see, for example, Fig.~7.5 of~\cite{moorethesis}). The perturbation of the outcome by an adversary is given by $\alpha = \delta + \omega$, the proportion of the $\nu$ rounds which have bits flipped. Let $\beta$ be the correct proportion of the $L-k$ bits used for parameter estimation which have the value 0:
\begin{equation}
    \beta = \frac{1}{2}\left(1 + \cos(\hat{\theta})\right)
\end{equation}
(recall that we are using the shorthand $\hat{\theta} := \hat{\overline{\theta}}_{\p}$). The order of the bits (i.e. which are 0 and which are 1) is random, and hence the new expected proportion is:
\begin{equation}
    \begin{split}
    \beta' &= \beta(1 - \alpha) + (1 - \beta)\alpha = \beta + \alpha - 2\alpha \beta \\
    &= \frac{1}{2}\left(1 + \cos(\hat{\theta}')\right).
    \end{split}
\end{equation}

We set $\eta := \hat{\theta}' - \hat{\theta}$, and then we can find:
\begin{equation} \label{eq: displaced estimate}
    \begin{split}
    |\alpha| &= \left|\sin(\hat{\theta} + \nicefrac{\eta}{2})\sin(\nicefrac{\eta}{2})/\cos(\hat{\theta})\right|.
    \end{split}
\end{equation}
We now aim to show that if $|\alpha|$ is sufficiently small, then $|\eta|$ is also small. 

We can lower bound the size of $\alpha$ according to $\eta$ and $\hat{\theta}$ (using truncated Taylor expansions):
\begin{equation}
    \begin{split}
        |\alpha| &= \left| \sin(\eta/2) \left( \tan(\hat{\theta}) \cos(\eta/2) + \sin(\eta/2)\right)\right| \\
        & \geqslant \frac{1}{2}|\tan(\hat{\theta}) \sin({\eta})| \\
        & \geqslant \frac{1}{2} \left| \hat{\theta} + \frac{\hat{\theta}^3}{3} + \frac{2 \hat{\theta}^5}{15} \right| \left| \frac{\eta}{2}\right|.
    \end{split}
\end{equation}

That is, $\alpha$ is at least polynomial in $\eta$. Hence, it is not possible to achieve arbitrarily large bias without incurring a polynomial cost in the number of parameter estimation rounds affected (which we can bound).

\subsection{Uncertainty}

We now consider the uncertainty potentially introduced by adversaries. The variance of the estimation typically scales as $1/\nu$~\cite{bollinger1996optimal} (or $1/\nu n^2$ if the proportion is instead expressed as $\beta = \frac{1}{2}(1 + \cos(n\theta))$, using slightly different notation).

Recall that we have defined $\alpha$ to be some proportion of the bitstring produced by \PARESTshort{} that are flipped by malicious behaviour. For the binomial distribution we can use the result that the variance for a single trial is $\beta (1- \beta)$ in the original case, and:
\begin{equation}
    \begin{split}
    \Delta^2 \hat{\beta}' &= (\beta + \alpha - 2\alpha \beta)(1 - \beta - \alpha + 2\alpha \beta) \\
    & = \sin^2 (\hat{\theta})/4 - \alpha (1-\alpha) \cos^2 (\hat{\theta})
    \end{split}
\end{equation}
in the corrupted case.

Hence, we can use the error propagation:
\begin{equation}
\begin{split}
       \Delta \theta' &= \frac{d \hat{\theta}'}{d \beta'} \Delta \beta' 
       = \frac{2}{\sin(\hat{\theta}')} \Delta \beta' \\
       &= \sqrt{\frac{\sin^2(\hat{\theta})}{\sin^2(\hat{\theta}')} - 4\alpha (1-\alpha) \frac{\cos^2(\hat{\theta})}{\sin^2(\hat{\theta}')}}. \\
       \end{split}
\end{equation}
Using:
\begin{equation}
    \begin{split}
    \sin^2(\hat{\theta}') &= 1 - (1 - 2 \beta')^2 \\
    &= 1 - \cos^2(\hat{\theta})(1 - 2\alpha)^2,
    \end{split}
\end{equation}
we arrive at:
\begin{equation}
    \begin{split}
        \Delta \hat{\theta}' &= 1.
    \end{split}
\end{equation}

Note that this is for a single round, but over the $\nu$ rounds, we reintroduce the $1/\nu$ factor. Therefore, we have rederived the unperturbed variance, and we can see that:
\begin{equation}
    |\Delta^2 \hat{\theta}' - \Delta^2 \hat{\theta}| = 0.
\end{equation}

The fact that the proportion of perturbed rounds, $\alpha$, does not influence the variance, may be understood by considering that for a fixed $\alpha$, the impact of the malicious behaviour on the parameter estimation is well-behaved. There is no prior assignment of which rounds are 0 and which are 1, and therefore being able to reverse the output of random rounds has a fixed effect on the expectation value without necessarily introducing any additional noise, and simply adding a bias. Alternatively, this can be understood through an attack strategy: if an adversarial non-participant was able to add a $\pi/4$ rotation, and remain undetected by \PARVERshort{}, this would disturb a high proportion of the \PARESTshort{} bits but could be reformulated as the estimation of a new parameter with the same variance.

\subsection{Summary of integrity}

The condition on the integrity of parameter estimation is the bias:
\begin{equation}\label{eq: bias}
    \Pr(|\mathds{E}(\hat{\theta}') - \overline{\theta}_{\p}| \geq \eta ) \leq \exp(-2 (f(\eta, \hat{\theta})- \delta)^2 \frac{(L-k) k^2}{L(k + 1)})
\end{equation}
where:
\begin{equation}
    f(\eta, \hat{\theta}) = \frac{1}{2} \left| \hat{\theta} + \frac{\hat{\theta}^3}{3} + \frac{2 \hat{\theta}^5}{15} \right| \left| \frac{\eta}{2}\right|.
\end{equation}

This rests on several assumptions, in particular that non-participants have no information about which rounds are used for \PARVERshort{} and that a true \GHZtext{} state is distributed to the network. The first assumption depends on the \ACKAshort{} protocols used. If the \ACKAshort{} protocol allows $a$ bits of key to be leaked, this corresponds directly to $\alpha L$ bits which can be flipped without detection (as in~\Cref{eq: verification failure}). 

The quality of the \GHZtext{} state is assured through the verification protocol. Much like the analysis in~\Cref{sec: parest bitflips}, we note that a state which is able to pass the \STAVERshort{} rounds would also produce the correct outcome at \PARESTshort{}, and hence, if there is a failure rate of $a$ rounds, this can once again be interpreted as $\alpha L$ affected bits of \PARESTshort{}, although we note that this may have an outsized change to $\beta'$, as adversaries can in this case force an outcome, e.g. make sure that the outcome bit is 1 -- hence, we can follow the same argument but instead with $\beta' = \beta + \alpha$.

It is important to highlight the role of $\delta$ here: this can be seen as a known bias introduced in the parameter estimation. On the one hand, as $\delta$ increases in size, Alice's confidence in the bias bound expressed in Eq.~\ref{eq: bias} decreases. This can alternatively be expressed that there is already a bias $\eta'$ given by:
\begin{equation}
    |\delta| = \left| \sin(\eta'/2) \left( \tan(\hat{\theta}) \cos(\eta'/2) + \sin(\eta'/2)\right)\right|
\end{equation}
and the total bias can now be expressed as:
\begin{equation}\label{eq: bias 2}
    \Pr(|\mathds{E}(\hat{\theta}') - \overline{\theta}_{\p}| \geq \eta + \eta' ) \leq \exp(-2 f(\eta, \hat{\theta})^2 \frac{(L-k) k^2}{L(k + 1)}).
\end{equation}

On the other hand, assuming $\delta$ is sufficiently small, using the measured \PARVERshort{} proportion $\beta'$, and $\delta$, Alice can correct the error by calculating:
\begin{equation}
    \beta = \frac{\beta' - \delta}{1 - 2\delta}.
\end{equation}
The bias is then simply~\Cref{eq: bias}, with $\delta=0$. This correction procedure assumes that the noise that creates the error $\delta$ in \PARVERshort{}~is equally likely to cause a bit flip in both \PARVERshort{}~and \PARESTshort{}~rounds; as shown previously, this is the case for attacks by potential adversaries, but may be true for other sources of noise (such as in Alice's experimental apparatus). Therefore, this relies on the experimental noise being sufficiently low, or well-characterised, as to only consider noise from adversaries or dishonest non-participants.

\section{Privacy}
\label{appendix:privacy}

In this work, we define the notion of \emph{privacy} as the property of a given scheme to be executed while ensuring that sensitive information that belongs to individual agents remains inaccessible to unauthorised parties.

\subsection{Definition of privacy}

Formally, \emph{privacy} in the context of networked sensing can be defined as follows (see the diagram below for an example).\\

\begin{minipage}[]{0.65\textwidth}

\begin{definition}
\label{def:privacy}
Let $\dis = \{d_{1}, \dots, d_{\sizeofd}\}$ be a subset of $\sizeofd$ dishonest agents selected from $\{a_{1}, \dots, a_{\numberofagents}\}$, the complete set of agents in the network; and let $\tar~=~\dis^{c}~=~\{t_{1}, \dots, t_{\numberofagents-\sizeofd}\}$ be the \emph{target subset} of size $\numberofagents-\sizeofd$, given by the complement of set $\dis$.

Let $\{ \theta^{(d)}_{i} \}$ be the local parameter of agent $d_{i}~\in~\dis$, and $\{ \theta^{(t)}_{j} \}$ the parameter of a target agent $t_{j}~\in~\tar$.

A protocol is \textbf{private} if the quantum Fisher information $\QFIm$ that any subset of agents $\dis$ can extract about any function $f~=~f(\theta^{(t)}_{1}, \dots, \theta^{(t)}_{\numberofagents-\sizeofd})$ throughout the protocol is:
\begin{align}
     \QFIm(f| \rho_{\dis}) \leq \QFIm(f | \avgp, \theta^{(d)}_{1}, \theta^{(d)}_{2}, \dots, \theta^{(d)}_{\sizeofd}).
\end{align}
Here $\rho$ represents the global classical-quantum state produced by the protocol, encompassing the final distributed quantum state and any exchanged classical information, and $\rho_{\dis}$ is the partial state that is shared by the dishonest agents.
In other words, agents in $\dis$ do not get more information about $f(\theta^{(t)}_{1}, \dots, \theta^{(t)}_{\numberofagents-\sizeofd})$ than they already have from sharing the state $\rho$ and from knowing the local parameters $\{ \theta^{(d)}_{i}\}$ of agents in $\dis$, the public value $\avgp$, and any function $g(\avgp, \theta^{(d)}_{1}, \theta^{(d)}_{2}, \dots, \theta^{(d)}_{\sizeofd})$.
\end{definition}
\end{minipage}
\hfill
\begin{minipage}[]{0.3\textwidth}
    \centering
    \includegraphics[width=1.0\textwidth]{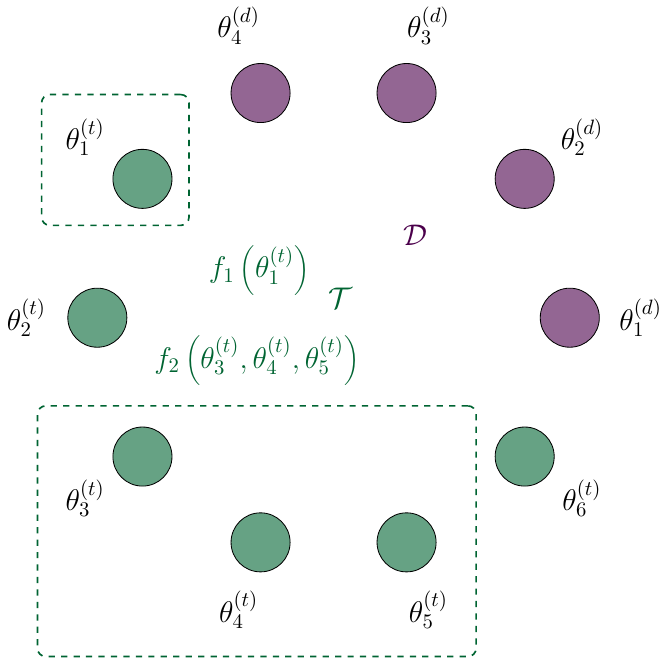}
    \label{fig:definition_privacy}
\end{minipage}

\subsection{Proof overview}

A quantifier of privacy in this scenario needs to capture the idea that, once the information of each local parameter $\theta_{i}$ is locally encoded in the state:
\begin{align}
    \rho = \frac{\ket{0}^{\otimes \numberofagents} + \ket{1}^{\otimes \numberofagents}}{\sqrt{2}}
    \qquad
    \longrightarrow
    \qquad
    \rho_{\vecp} = \frac{\ket{0}^{\otimes \numberofagents} + e^{i \avgp} \ket{1}^{\otimes \numberofagents}}{\sqrt{2}},
\label{eq:rho_theta}
\end{align}
only information about $\avgp = (\nicefrac{1}{\sizeofp})\sum_{i:a_{i} \in \p} \theta_{i}$ can be extracted from the system, but each secret parameter $\theta_i$ remains hidden. In our protocol, this guarantee essentially follows from the work of~\cite{shettell2022private, hassani2024privacy, Bugalho2025privaterobuststates}, as any step that involves parameter input uses the same states and operations as described in previous iterations of the protocol.

With this definition of privacy in mind, we can introduce $L_{i}$, the symmetric logarithmic derivative (SLD) for the parameter $\theta_{i}$. $L_{i}$ is a Hermitian operator given by the relation
\begin{align}
\label{eq:SLD}
    L_{i} \; \rho_{\vecp} +  \rho_{\vecp} \; L_{i} = 2 \; \frac{\partial \rho_{\vecp}}{\partial \theta_{i}}
\end{align}
and it allows us to compute the coefficients of the quantum Fisher information matrix (QFIm, $\QFIm$) \cite{helstrom1969quantum, Liu_2020}.

The QFIm is a symmetric matrix with real elements that quantifies the amount of extractable information about different unknown parameters over all possible measurements. Using the SLD, the elements of this matrix can be calculated as
\begin{align}
    \label{eq:QFIm}
    \QFIm_{ij}[\vecp]=\frac{1}{2} \Tr\left[\rho _{\vecp}\{L_{i}, L_{j}\}\right].
\end{align}
This means that the QFI for parameter $\theta_{i}$ is
\begin{align}
    \label{eq:QFIm_ii}
    \QFIm_{ii}[\vecp] = \Tr\left[\rho _{\vecp} L_{i}^2\right]
\end{align}
from which we can see that the presence of non-zero off-diagonal entries of the QFIm implies statistical correlation between the local parameters, as there is a way of extracting information of $\theta_{i}$ and $\theta_{j}$ simultaneously. Equivalently, if the different SLDs do not commute, the different parameters cannot be estimated independently. On the other hand, if the QFIm is block diagonal, parameters in different blocks are \textit{information orthogonal}, in the sense that their maximum likelihood estimates are asymptotically uncorrelated.

As the aim is to estimate one function of the unknown parameters, $\avgp =f(\vecp)$, the corresponding QFIm may be obtained  by reparametrisation, that is:
\begin{align*}
\QFIm' [\vecp] &= B^{T} \QFIm [\vecp] B
\end{align*}
where $B$ is a transformation matrix into an orthogonal basis such that the first element is the desired linear combination. Privacy (as defined in this work) is ensured if $\QFIm'$ is a rank-1 matrix, as this implies that only information about one linear combination of the parameters can be retrieved: the mean of all the local parameters.

Following the derivation from \cite{hassani2024privacy}, the linear function of interest can be encoded in a vector $\mathbf{w} = (\omega_{1}, \dots, \omega_{\sizeofp})^{T}$ (cf. \cite{proctor2017networked}), so that:
\begin{align}
    \label{eq:w}
    \avgp = \mathbf{w}^{T} \vecp =
    \left( \omega_{1}, \dots, \omega_{\sizeofp} \right)
    \begin{pmatrix}
    \theta_{1} \\
    \vdots \\
    \theta_{\sizeofp}
    \end{pmatrix}.
\end{align}
As we are only interested in the function given by $\mathbf{w}$, we can choose the matrix:
\begin{align}
    \label{eq:W-linear}
    W \equiv \mathbf{w}\mathbf{w}^{T} &=\left(\begin{array}{cccc}
    \omega _{1}\omega _{1}&\omega _{1}\omega _{2}&\cdots&\omega _{1}\omega _{\sizeofp}\\ 
    \omega _{2}\omega _{1}&\omega _{2}\omega _{2}&\cdots&\omega _{2}\omega _{\sizeofp}\\
    \vdots&\vdots&\ddots&\vdots\\
    \omega _{\sizeofp}\omega _{1}&\omega _{\sizeofp}\omega _{2}&\cdots&\omega _{\sizeofp}\omega _{\sizeofp}\\
    \end{array}\right),
\end{align}
and then $Q' \propto W$ can be used for the desired purpose of this work, as it is a rank-1 matrix that carries the information of the average value. This implies that:
\begin{align}
    \QFIm_{ij} \propto \omega_i \omega_j \quad \forall i,j.
\end{align}

Using the continuity relation of the quantum Fisher information, as shown in detail in~\cite{hassani2024privacy}, the following condition must be satisfied:
\begin{equation}\label{eq:Priv-Condition-linear}
    \Vert\partial _{i}\rho _{\vecp}-\partial _{j}\rho _{\vecp}\Vert_{\mathrm{tr}}\propto\vert\omega _{i}-\omega _{j}\vert,~~\forall i \neq j.
\end{equation}
In this particular case, the aim is to compute the average value of the local parameters:
\begin{align}
    \label{eq:w_avg}
    \avgp = \mathbf{w}^{T} \vecp =
    \left( \frac{1}{m}, \dots, \frac{1}{m} \right)
    \begin{pmatrix}
    \theta_{1} \\
    \vdots \\
    \theta_{m}
    \end{pmatrix}
\end{align}
which means that Eq. \eqref{eq:W-linear} becomes:
\begin{align}
    \label{eq:W-linear-mean}
    \mathbf{w}\mathbf{w}^{T} &= \frac{1}{\sizeofp^{2}} \left(\begin{array}{cccc}
    1 & 1 & \cdots & 1\\
    1 & 1 & \cdots & 1\\
    \vdots&\vdots&\ddots&\vdots\\
    1 & 1 & \cdots & 1\\
    \end{array}\right)
\end{align}
and, since $\omega_i = \omega_j = \frac{1}{m}$ for all $i,j$, Eq. \eqref{eq:Priv-Condition-linear} becomes simply:
\begin{align}
    \label{eq:condition-average}
    \partial_{i} \rho_{\vecp} = \partial_{j} \rho_{\vecp}, \qquad \forall i,j.
\end{align}
Therefore, any state that satisfies Eq. \eqref{eq:condition-average} can be used for the private estimation of the average value.

Using this,~\cite{hassani2024privacy} shows that for an $\numberofagents$-partite \GHZtext{} state and encoding unitaries $\mathbf{\Lambda}(\vecp) = \bigotimes _{i=1}^{\numberofagents}(\ketbra{0}{0} + e^{i \frac{\theta_{i}}{m}}\ketbra{1}{1})$,  the only information that can be retrieved from the resulting state:
\begin{equation}
    \GHZstatetheta=\frac{1}{\sqrt{2}}\left(\ket{0}^{\otimes \numberofagents}+ e ^{i \nicefrac{(\theta_1 + \dots + \theta_\numberofagents)}{m}}\ket{1}^{\otimes \numberofagents}\right)
\end{equation}
corresponds to the sum of all the local parameters, and therefore the local parameters of users remain private.

More generally, for an ideal state $\sigma$, any state $\rho$ that originates from the output of any state verification scheme that ensures 
\begin{align}
\Pr\left(\|\sigma - \rho\|_{\mathrm{tr}} \geqslant \varepsilon_{\STAVERshort{}}\right) \leqslant \alpha, 
\end{align}
for some security parameter $\varepsilon_{\STAVERshort{}}$ and some confidence bound $\alpha$, which can be made arbitrarily small. From \cite{hassani2024privacy}, we know that the `$\varepsilon$-privacy' (a variable that quantifies the leakage of information about the local parameters $\theta_i$) can be bounded with probability $1-\alpha$ as:
\begin{align}
    \varepsilon_{\mathrm{priv}} \leq 4 \; ||M||_{\infty} || \sigma - \rho ||_{\text{tr}} \leq 4 \; ||M||_{\infty} \; \varepsilon_{\STAVERshort{}} 
\end{align}
where $M$ is an operator that contains the information about the local encoding of the parameters. For the case of the average value of the local parameters, $||M||_\infty=||\sigma_z/2||_{\infty}=\frac{1}{2}$.

A similar definition of $\varepsilon$-privacy is defined in\cite{Bugalho2025privaterobuststates}.

\section{Anonymity}
\label{appendix:anonymity}
We make use of the definition of \emph{full anonymity} from \cite{grasselli2022secure} and adapt it to our setting. 
This definition compares the output state of the protocol against an \emph{ideally anonymous} state, which is any state that is perfectly anonymous under a specific requirement, introduced below.
Note that there is not one unique ideally anonymous state, but that any state which adheres to the requirement will be ideally anonymous. We first introduce the structure of the output state, and state the requirement for the output state to be ideally anonymous.
Utilising the concept of ideally anonymous states, we can define anonymity of an \APPEshort{} protocol, which is in terms of closeness to any such ideally anonymous state.
Subsequently, we prove our protocol's anonymity under this definition.

The anonymity statement in \Cref{sec:prop-anon} presents our results with respect to a confidence window inherited from the preceding \STAVERshort{} protocol. In the following however we omit the confidence window, essentially taking $\alpha = 0$. 

\subsection*{Ideally anonymous states and definition of full anonymity}
The output state $\sout[\reg]$ is defined on the registers $P$, $T$, $K$, $C$ and $E$, all containing classical information. 
Each of these registers can be indexed by the agents, so that e.g.~$T_{t} = 0.01$ indicates that agent $a_{t}$ holds the value $0.01$ in their part of the register $T$. 
Note that an agent can only access the entries of the registers at their own index. Within this section, let $i$ be the specific index such that $a_{i} = \alice{}$, and let $\vec{j}$ be a vector with $1$ at indices corresponding to participants and $0$ otherwise. Using this notation, the registers are defined as:
\begin{itemize}
    \item[$P$:] This register contains the information regarding the roles in the network for every agent. It holds that:
    \begin{equation*}
        P_t=\begin{cases}
       \vec{j} & \text{if $t=i$}\\
            1 & \text{if $a_t\in\p \setminus \alice$,}\\
            0 & \text{otherwise.}
        \end{cases}
    \end{equation*}
    \item[$T$: ] This register holds the parameters of the agents: $T_{t}=\theta_{t}$.
    \item[$K$: ] This register contains the secret key $\kappa$ for the participants:
   \begin{equation*}
        K_{t}=\begin{cases}
            \kappa & \text{if } a_{t} \in \p,\\
             \emptyset & \text{otherwise.}\\
       \end{cases}
    \end{equation*}
    \item[$C$: ] This register is divided into three sub-registers, defined as follows:
    \begin{itemize}
        \item $C_{NV}$ contains all (classical) information from \NOTIFICATION{} and \VOTE,
        \item $C_{SV}$ contains all information from \STAVERshort{}, and
        \item $C_{PP}$ contains all public classical communication from \PARVERshort{} and \PARESTshort{}.
    \end{itemize} 
    \item[$E$: ] This register contains all classical and quantum information held by the adversary Eve. This may include the information of one or more dishonest nodes.
\end{itemize}
Note that we have omitted Alice's measurement outcome and $\avgp$ from being included in any of the registers to ease our analysis. While these are crucial for integrity and privacy, they are irrelevant for anonymity, because Alice never announces her measurement outcome.

The output state $\sout$ depends on Alice and her choice of participants, as well as on the set of dishonest parties, $\dis \subsetneq \{1,\dots,\numberofagents{}\}$. We capture this by using the following notation:
\begin{equation*}
    \soutdis[\reg|i,\vec{j}].
\end{equation*}

This notation allows us to precisely state our requirement for ideally anonymous states. Even though the output state $\sout$ is dependent on Alice and her choice of participants, we define a state to be anonymous if the reduced state of $\sout$ on any relevant subset is independent of Alice and the participants.
More specifically, let $\g\subseteq\{1,\dots,n\}$ be a subset of agents, and $\g^c = \{1,\dots,n\} \setminus \g$ be its complement. 
For a given register $S$, we define $S_\g=\{S_i\in S| i\in \g\}$, and for a state $\rho_S$, we have $\rho_{S_\g}=\Tr_{S_{\g^c}}(\rho_S)$. 
Using this notation, we can now precisely state the requirement for $\soutdis[\reg]$ to be ideally anonymous.

\begin{definition}\label{def:anonymity}
Let $\dis$ be the set of dishonest agents in an \APPEshort{} protocol with an output state $\soutdis[\reg]$. The state $\soutdis[\reg]$ is \emph{fully ideally anonymous with respect to a subset} $\g \subseteq\{1,\dots,n\}$ if it holds that:
\begin{equation}\label{eq:anonymity}
    \soutdis[\regg|i,\vec{j}] =  \soutdis[\regg|i',\vec{j'}],
\end{equation}
for any $i, i', \vec{j},\vec{j'}$ such that $i,i'\notin\g\cup\dis$, $\vec{j}\cap\dis=\vec{j'}\cap\dis$ and $\vec{j}\cap\g=\vec{j'}\cap\g$. Moreover, because we specifically fix the size of the set of participants, it should hold that $H(\vec{j}) = H(\vec{j'})$, where $H(\vec{j}) = \sum_{k} j_{k}$ denotes the \emph{Hamming weight} of $\vec{j}$, i.e.~its total number of 1's.

This state $\soutdis[\reg]$ is \emph{fully ideally anonymous} with respect to the set of dishonest agents $\dis{}$ if it is fully ideally anonymous with respect to every subset $\g \subseteq\{1,\dots,n\}$.
\end{definition}

With the definition of a fully ideally anonymous state, we are now equipped to define $\varepsilon_a$-\emph{full anonymity} of our \APPEshort{} protocol. 

\begin{definition}\label{def:epsanonymity}
    An \APPEshort{} protocol with an output state $\rodis[\reg]$ is \emph{$\varepsilon_{a}$-fully anonymous} with respect to a set of dishonest agents $\dis{}$ if there exists a fully ideally anonymous state $\soutdis[\reg]$ such that:
    \begin{equation}
        \| \rodis[\reg] - \soutdis[\reg] \|_{\mathrm{tr}} \leqslant \varepsilon_{a},
    \end{equation}
    where $\|\cdot \|_{\mathrm{tr}}$ denotes the trace distance.
\end{definition}

Note that, similar to security in quantum key distribution (QKD) protocols \cite{portmannCryptographicSecurityQuantum2014,tomamichel2017largely}, we have defined anonymity with respect to an anonymity parameter $\varepsilon_a$. This captures the notion of \emph{quantifiable} anonymity, which allows for the protocol to be fully anonymous in non-perfect scenarios, e.g.~due to noise. 
One can then obtain anonymity \emph{approximately}, meaning that the anonymity statements hold except for increasingly small probabilities.
In the following subsections, we prove that our protocol satisfies \Cref{def:epsanonymity}.

\subsection*{Proof overview}
In order to show that our \APPEshort{} protocol is $\varepsilon_{a}$-fully anonymous, we show that its output state $\rodis[\reg]$ is $\varepsilon_{a}$-close to a particular fully ideally anonymous output state $\soutdis[\reg]$.

This specific output state is defined as the output of \APPEshort{} in a completely ideal scenario; we can implicitly model our protocol as a CPTP map $\prch$, so that $\soutdis[\reg] = \prch(\si)$, where $\si$ is an idealised input state to our protocol.

More specifically, $\si = \si[R,I,E]$ is defined over three registers. $R$ contains the quantum states used in the protocol, $I$ the other inputs to the protocol (i.e.~the index of Alice, her choice of participants, and the parameters of all the agents), and $E$ Eve's quantum and classical side information.

We first observe that $\STAVERshort{}$ is completely independent of the roles of the agents, and therefore we can ignore this step in our proof. This means that the register $R$ can be regarded to only hold the verified GHZ states, so that we can write $\tau_{R,I,E}$ as (recalling that $L = \nu + k$ is the total number of verified \GHZtext{} states used for $\PARESTshort{}$ and $\PARVERshort{}$):
\begin{equation}\label{eq:input_state}
    \si=\GHZstate\GHZstatebra_{\quantreg}^{\otimes L} \otimes \sigma_{I} \otimes \sigma_E.
\end{equation}

Furthermore, \STAVERshort{}  guarantees that for some $\varepsilon_{\mathrm{ver}}>0$ it holds that $\|\rho^{\mathrm{in}}_{R}-\sigma^{\mathrm{in}}_{R}\|_{\mathrm{tr}} \leqslant \varepsilon_{\mathrm{ver}}$, where $\ri$ is the actual input state of the protocol - note that, because $\sigma^{\mathrm{in}}_{R}$ is pure, it can be concluded that $\rho^{\mathrm{in}} = \rho^{\mathrm{in}}_{R} \otimes \rho^{\mathrm{in}}_{I,E}$. As the register $I$ does not change between the ideal and real case, it holds that $\rho^{\mathrm{in}}_{I} = \sigma^{\mathrm{in}}_{I}$. Finally, all communication after \STAVERshort{} is only correlated with the register $R$, so there is no loss of generality in assuming that $\rho^{\mathrm{in}}_{E} = \sigma^{\mathrm{in}}_{E}$, so that it holds that $\|\rho^{\mathrm{in}}-\sigma^{\mathrm{in}}\|_{\mathrm{tr}} \leqslant \varepsilon_{\mathrm{ver}}$.

We then have $\rodis[\reg] = \prch(\rho^{\mathrm{in}})$ and can leverage the contractivity of the trace distance under quantum channels (data processing inequality) to obtain:
\begin{equation}\label{eq:proofofepsanon}
\|\ro - \sigma^{\dis}\|_{\mathrm{tr}} = 
\|\prch(\ri)-\prch(\si)
\|_{\mathrm{tr}} \leqslant \|\ri-\si\|_{\mathrm{tr}} \leqslant \varepsilon_{\mathrm{ver}}.
\end{equation}
Proving anonymity of our \APPEshort{} protocol therefore reduces to showing that $\soutdis[\reg] = \prch \left(\si\right)$ is a fully ideally anonymous state under \Cref{def:anonymity}. Indeed, from \eqref{eq:proofofepsanon} it then follows that our \APPEshort{} protocol is $\varepsilon_{\mathrm{ver}}$-fully anonymous under \Cref{def:epsanonymity}.

In the remainder of this section we prove that $\soutdis[\reg]$ is a fully ideally anonymous state, first in the honest ($\dis=\emptyset$), and later in the dishonest setting (i.e.~for arbitrary $\dis$, as long as $\alice{} \not \in \dis$).

\subsection*{\texorpdfstring{$\soutdis[\reg]$}{Output state} is fully ideally anonymous in the honest setting}
Our goal is now to show that $\soutdis[\reg]$ is fully ideally anonymous under \Cref{def:anonymity} in the honest setting, i.e.~where $\dis = \emptyset$, allowing us to write $\soutdis[\reg] = \sout[\reg]$. 

As a starting point, we note that although there exists correlations between various registers $P,\ T,\ K,\ C$ and $E$ the ideal state is separable over a certain partitioning of the registers, which allows us to write $\sout[\reg]$ in tensor product form:
\begin{equation}\label{eq:sigmainsepform}
\sout[\reg] = \sigma_{PTKC_{NV}}\otimes\sigma_{C_{PP}}\otimes\sigma_{C_{SV}}\otimes\sigma_E.
\end{equation}

The registers $P$, $T$, $K$ and $C_{NV}$ cannot be written in tensor product form. Indeed, only the participants obtain the secret key so that $P$ and $K$ are correlated, similarly only the participants keep their parameter in their register $T$, and \NOTIFICATION{} and \VOTE{} correlate with the contents of $P$ as well.
An important part of our later analysis shows that $C_{PP}$ is not correlated with any other register.
The tensor product structure of \eqref{eq:sigmainsepform} then follows from the fact \STAVERshort{} is independent of the rest of the protocol, so that $\sigma_{C_{SV}}$ is separate from the rest of the output state.
Finally, there is no loss of generality in assuming that $\sigma_{E}$ is uncorrelated, because all relevant (side)-information (e.g.~the public communication) can be understood to be explicitly considered already within the other registers.

To prove that $\sout[\reg]$ is fully ideally anonymous under \Cref{def:anonymity}, it suffices to show that the tensor factors $\sigma_{PTKC_{NV}}$, $\sigma_{C_{PP}}$, $\sigma_{C_{SV}}$ and $\sigma_E$ individually satisfy \cref{eq:anonymity}.

\begin{itemize}
    \item[$\sigma_{E}$] \NOTIFICATION{} and \VOTE{} rely on private pairwise communication, therefore Eve cannot learn anything from these protocols that was not shared publicly. Moreover, the input state $\tau_{R,I,E}$ is separable between all three registers (by \eqref{eq:input_state}), so that the $\sigma_{E}$ remains completely separable from the networks' registers, and the identities of the agents. Hence, $\sigma_{E}$ trivially satisfies \eqref{eq:anonymity}.
    \item[$\sigma_{C_{SV}}$]As noted before, there is no distinction between participants and non-participants in \STAVERshort{}, i.e.~every agent in the network performs the exact same steps, regardless of their role. This means that $\sigma_{C_{SV}}$ trivially satisfies \eqref{eq:anonymity}.
    \item[$\sigma_{PTKC_{NV}}$] By construction, the registers $P$, $T$ and $K$ together obey \eqref{eq:anonymity} for any subset $\mathcal{G}$ and any choice of $i,i'$ and $\vec{j},\vec{j'}$. From \cite{broadbent2007information} we know that both \NOTIFICATION{} and \VOTE{} are anonymous, so that the classical communication involved with the protocols are independent of Alice and her choice of participants. However, \VOTE{} outputs $\sizeofp{}$, which makes $C_{NV}$ dependent on $P$. Nevertheless, \Cref{def:anonymity} specifically only compares states with the same number of participants (i.e.~$H(\vec{j}) = H(\vec{j'})$), so that the requirement \eqref{eq:anonymity} is still met. It therefore holds that $\sigma_{PTKC_{NV}}$ is fully ideally anonymous under \Cref{def:anonymity}.
    
    \item[$\sigma_{C_{PP}}$ ] To show that $\sigma_{C_{PP}}$ satisfies \eqref{eq:anonymity}, we prove that its contents are uniformly random by virtue of Alice announcing a random bit instead of her actual measurement outcome.
    
   Apart from Alice's random bit, $\sigma_{C_{PP}}$ contains the announced outcomes of the $X$-basis measurement on $\GHZstate = (\nicefrac{1}{\sqrt{2}})\left(\ket{0}^{\otimes n}+\ket{1}^{\otimes n}\right)$ in the \PARVERshort{} rounds and $\GHZstatetheta = (\nicefrac{1}{\sqrt{2}})\left(\ket{0}^{\otimes n} + e^{i\avgp} \ket{1}^{\otimes n}\right)$ in the \PARESTshort{} rounds.  
   While the outcomes of the $X$-basis measurements obey parity statistics for both \PARESTshort{} and \PARVERshort{} rounds, this is only the case when Alice's outcome is also considered. However Alice announces a random bit, ensuring that the \emph{announced} measurement outcomes stored in $C_{PP}$ are uniformly random and uncorrelated, and thus independent of the choice of $\p$. 
   
   To this end we show that for any strict subset it holds that all measurement outcomes are equally likely. Indeed, consider $A \subsetneq \{1,2,\dots,n\}$, we can then write the reduced states on $A$ as:
    \begin{equation}
        \Tr_{A^{c}}(\GHZstate\GHZstatebra)=\Tr_{A^{c}}(\GHZstatetheta\GHZstatethetabra)=\frac{\ketbra{0}{0}^{\otimes \abs{A}}+\ketbra{1}{1}^{\otimes \abs{A}}}{2}.
    \end{equation}

Consider now the probability $\Pr\left(\{o_{j}\}_{j \in A}\right)$ that the agents in $A$ obtain a specific set of outcomes $\{o_{j}\}_{j \in A}\in\{0,1\}^{|A|}$ for their $X$-basis measurements.
A straightforward calculation reveals that the probability of obtaining this specific outcome $\{o_{j}\}_{j \in A}$ is $\Pr\left(\{o_{j}\}_{j \in A}\right) =\nicefrac{1}{2^{|A|}}$.
That is, each outcome is equally likely.
As this holds for any proper subset $A$, it follows that the partial measurement outcomes are uniformly random and uncorrelated. It follows that the contents of $C_{PP}$ are independent of the choice of $\p$, and therefore that $\sigma_{C_{PP}}$ obeys \eqref{eq:anonymity}.
\end{itemize}

Finally, we note that due to the fact that the correlations in the measurement outcomes are only between the complete set of outcomes, and that any subset of measurement outcomes is uniformly random and uncorrelated, it is actually not necessary for Alice to announce a random bit instead of her actual measurement outcome to safeguard anonymity. However, the tensor product structure of \eqref{eq:sigmainsepform} is then lost, as $C_{PP}$ will depend on $T$ and $P$, making the notation less clear. Furthermore, as briefly discussed in \Cref{sec:discussion}, this would introduce the need of a simultaneous broadcast channel, or an adaptation to the protocol so that the independence of all measurement outcome announcements can be guaranteed.

We proved that $\sigma_{PTKC_{NV}}$, $\sigma_{C_{PP}}$, $\sigma_{C_{SV}}$ and $\sigma_E$ all satisfy \eqref{eq:anonymity} and therefore by \eqref{eq:sigmainsepform} we know that $\sout[\reg]$ obeys \Cref{def:anonymity} and thus our protocol is fully anonymous in the honest setting, where $\dis=\emptyset$.

\subsection*{\texorpdfstring{$\soutdis[\reg]$}{Output state} is fully ideally anonymous in the dishonest setting}

We now turn our attention to the case where $\dis$ is non-empty, i.e.~there are nodes that are deviating from the protocol to try and learn the roles of other agents of the network. Importantly, the dishonest agents may base their strategy on the announcements of the honest agents.
Indeed, while the honest agents announce their measurement results for both \PARVERshort{} and \PARESTshort{}, the dishonest members can delay their measurement and announce anything else instead. 
The intermediate output state relevant to our analysis, which is the state \emph{before} the dishonest agents perform their measurement, is therefore:
\begin{equation}\label{eq:intermediatedishoneststate}
    \sigma_{PTKC_{NV}} \otimes \sigma_{SV} \otimes \rho_{\tilde{C}_{PP}R_{\dis}} \otimes \sigma_{E},
\end{equation}
where, $\rho_{\tilde{C}_{PP}R_{\dis}}$ is the classical-quantum state on the classical register $\tilde{C}_{PP}$ and the quantum register $R_{\dis}$. We write $\tilde{C}_{PP}$ to emphasize that the contents of the register differ from the honest case, because the dishonest agents have not measured their qubits and announced something else. ${R_{\dis}}$ is the part of the quantum register that has not been measured, i.e.~the qubits of the agents in $\dis$. Ultimately, the output state $\soutdis[\reg]$ will be obtained from the state in \eqref{eq:intermediatedishoneststate}.

To show that the output state is fully ideally anonymous it therefore suffices to show that the state in \eqref{eq:intermediatedishoneststate} is itself fully ideally anonymous (similar to \Cref{def:anonymity} but adapted to reflect the change of registers).
We can again exploit the tensor product structure and analyse the tensor factors individually; neither the states $\sigma_{PTKC_{NV}}$ nor $\sigma_{SV}$ have changed from the honest setting, which means that they are still fully ideally anonymous.

Although the output registers $C_{PP}$ and $E$ may become correlated by any deviation of the protocol by the dishonest agents, the intermediate state of \eqref{eq:intermediatedishoneststate} indeed does allow the tensor product structure, so that we may analyse $\rho_{\tilde{C}_{PP}R_{\dis}}$ separately from $\sigma_{E}$. In general, the state $\rho_{\tilde{C}_{PP}R_{\dis}}$ is a CQ-state that correlates any possible set of measurement outcomes $\{o_{j}\}_{\hon}$ of the honest agents with a state $\rho^{\dis}_{R_{\dis}|\{o_{j}\}_{\hon}}$ on $R_{\dis}$, which, in turn, can be computed as:
\begin{equation}\label{eq:reduced_after_announced}
     \rho^{\dis}_{R_{\dis}|\{o_{j}\}_{\hon}} = \left(\bra{\{o_{j}\}_{\hon}} \otimes I_{\dis} \right) \GHZstatethetaketbra \left(\ket{\{o_{j}\}_{\hon}} \otimes I_{\dis} \right).
 \end{equation}
 
In order to prove that $\rho^{\dis}_{R_{\dis}|\{o_{j}\}_{\hon}}$ is fully ideally anonymous, we now show that it is independent of the choice of participants.
To this effect we first rewrite the \GHZtext{} state as:
\begin{equation}
    \GHZstatetheta=\frac{1}{\sqrt{2^{n+1}}}\left(
    \left(1+\e^{i\avgp}\right)\sum_{\Delta\left(\{o_{j}\}\right) = 0}\ket{\{o_j\}}+  \left(1-\e^{i\avgp}\right)\sum_{\Delta\left(\{o_{j}\}\right) = 1}
    \ket{\{o_j\}}\right),
\end{equation}
where $\Delta_{\{o_{i}\}}$ denotes the parity of $\{o_{i}\}$, i.e.~its Hamming weight modulo two. From this it follows:
\begin{equation}
\begin{aligned}
    \left(\bra{\{o_{j}\}_{\hon}} \otimes I_{\dis} \right) \GHZstatetheta
    =\frac{1}{\sqrt{2^{|\dis|+1}}}\Bigg(
    &\left(1+\e^{i\avgp}\right)\sum_{\Delta_{\{o_{j}\}_{\dis}} =\phantom{1-} \Delta_{\{o_{j}\}_{\hon}}}\ket{\{o_j\}_{\dis}}\\+  &\left(1-\e^{i\avgp}\right)\sum_{\Delta_{\{o_{j}\}_{\dis}} =1- \Delta_{\{o_{j}\}_{\hon}}}
    \ket{\{o_j\}_{\dis}}\Bigg).
\end{aligned}
\end{equation}

It is directly apparent that this state is independent of the actual choice of participants but rather depends only on $\avgp$ which is irrelevant for anonymity. There is therefore no information on the roles of the agents that dishonest members can gain by deviating from the protocol. Finally we can conclude that the protocol is fully anonymous under \Cref{def:epsanonymity} in the dishonest setting as well.

\section{Sub-protocols}
\label{appendix:sub-protocols}

In this section, we present classical sub-protocols used throughout our work, with adapted notation.
The \APPEshort{} protocol presented in this work makes use of two classical subprotocols: \NOTIFICATION{} and \VOTE{}. Although these steps can be implemented in any way that preserves the anonymity of the participants (see \Cref{appendix:anonymity}), a particular choice of algorithms for these initial steps is presented by Broadbent \& Tapp \cite{broadbent2007information}.

In the \NOTIFICATION{} protocol, a coordinating agent is able to notify all participating agents while preserving the anonymity of every party involved. The output, which every agent computes locally and in private, reveals to each agents whether they participate or not in the overall \APPEshort{} protocol, and no other information can be extracted from it. This subprotocol, with adapted notation, is presented as follows.

\begin{protocolfloat}[h!]
\begin{protocol}{\NOTIFICATION{} \hfill (Broadbent \& Tapp 2007)}{A coordinator Alice ($\alice$), a set of $\sizeofp$ participants $\p \subseteq \{a_1, \dots, a_\numberofagents\}$.}{$\alice$~notifies the $\sizeofp$ participants.}
    \begin{protocollist}
        \item Each agent $\{ a_{i}\}_{i=1}^{\numberofagents}$ sends every other agent $\{ a_{j}\}_{j=1}^{\numberofagents}$, $\numberofagents$ random bits $\{ r_{ijk}\}_{k=1}^{\numberofagents}$, such that:
                \begin{itemize}
            \item if $a_{i} = \alice$, the random bits satisfy $\bigoplus_{j=1}^{\numberofagents} r_{ijk} =
                \begin{cases}
                    1, & \text{if $a_{k} \in \p$}\\
                    0, & \text{if $a_{k} \in \np$}
                \end{cases}$

            \item if $a_{i} \neq \alice$, the random bits satisfy $\bigoplus_{j=1}^{\numberofagents} r_{ijk} = 0$

            \end{itemize}
                    \item Each agent $\{a_{j}\}_{j=1}^{\numberofagents}$ receives the bits $\{r_{ijk}\}_{i=1}^{\numberofagents}$, computes $t_{jk} = \bigoplus_{i=1}^{\numberofagents} r_{ijk}$

        and sends $\{t_{jk}\}_{k=1}^{\numberofagents}$ to agent $a_{k}$.

        \item Each agent $\{a_{k}\}_{k=1}^{\numberofagents}$ receives $\{t_{jk}\}_{j=1}^{\numberofagents}$, and computes:
        \begin{equation}
             z_{k} = \bigoplus_{j=1}^{\numberofagents} t_{jk}
             = 
                 \begin{cases}
                 1, & \text{if $a_{k} \in \p$}\\
                     0, & \text{if $a_{k} \in \np$}
                 \end{cases}
        \end{equation}
    \end{protocollist}
    \label{prot:notification}
\end{protocol}
\end{protocolfloat}

For the second stage of the \APPEshort{} protocol, two subprotocols are needed: \PARITY{} and \VOTE{}. In the first one,
all agents are able to input a binary value, and all agents can compute the global parity of the network. This is used as a subroutine in the \VOTE{} protocol. In this subprotocol, agents anonymously input their \emph{choice} (in this case, 0 or 1, depending on whether they are participants or non-participants). At the end of this stage, each agent is able to compute the number of participants in the scheme, and decide whether to continue with the protocol or not. The \PARITY{} and \VOTE{} protocols are presented below, with adapted notation. The probability of an incorrect outcome is exponentially small in the number of rounds, $s$, and simultaneous broadcast is required to ensure that corrupt participants do not receive partial information which could be used in future rounds, potentially compromising the anonymity of the scheme.

\begin{protocolfloat}[h!]
\begin{protocol}{\PARITY{} \hfill (Broadbent \& Tapp 2007)}{Parities $\{x_i\}$, with each $x_{i} \in \{0, 1\}$.}{Each agent computes the global parity $y = x_{1} \oplus x_{2} \oplus \dots \oplus x_{\numberofagents}$.}

    Each agent $a_{i}$ does the following:
    \begin{protocollist}
        \item Select uniformly at random an $\numberofagents$-bit string $r_{i}~=~r_{i}^{(1)} r_{i}^{(2)} \dots r_{i}^{(\numberofagents)}$ with Hamming weight of parity $x_{i}$.
        \item Send $r_{i}^{(j)}$ to agent $a_{j}$ using the private channel; keep bit $r_{i}^{(i)}$ to themselves.
        \item Compute $z_{i}$, the parity of all the bits received, including $r_{i}^{(i)}$.
        \item Use the simultaneous broadcast channel to announce $z_{i}$.
        \item After the simultaneous broadcast is finished, compute $y = \oplus_{j=1}^{n} z_{j}$, the outcome of the protocol. If the simultaneous broadcast fails, the protocol aborts.

    \end{protocollist}
    \label{prot:parity}
\end{protocol}
\end{protocolfloat}

\begin{protocolfloat}[h!] \begin{protocol}{\VOTE{} \hfill (Broadbent \& Tapp 2007)}{Choice $x_{i} \in \{ 0, 1 \}$ voted by each agent ($x_i=1$ if $a_i \in \p$, $x_i=0$ if $a_i \in \np$), the number of rounds $s$.}{Each agent computes the tally $y = (y[0],y[1]) = (\numberofagents-\sizeofp, \sizeofp)$ with the number of votes for each candidate (i.e. the number of non-participants, and the number of participants, respectively).}

For each choice $b \in \{0, 1\}$:

\textbf{Phase A}:\\
        For each round $j \in \{1, \dots, s\}$:
            
        \begin{protocollist}
            \item each agent $a_{i}$ sets the value of $p_{i}$ in the following way: if $x_{i} \neq b$, $p_{i} = 0$; otherwise, $p_i=1$ with probability $\frac{1}{\numberofagents}$ and $p_i=0$ with complementary probability.
            \item the participants execute the \PARITY~protocol to compute the parity of $p_{1}, p_{2}, \dots, p_{n}$, but instead of broadcasting their output bit $z_{i}$, they store it as $z[b]_{i}^{j}$.
        \end{protocollist}
        \textbf{Phase B}:\\
        All agents $\{ a_{i}\}$ simultaneously broadcast $z[b]_{i}^{j}$ ($j \in \{1, \dots, s\}$). If the simultaneous broadcast is not successful, the protocol aborts.

    \textbf{Phase C}:\\
        To compute the tally $y[b]$, each participant sets: $p[b]_{j} = \bigoplus_{i=1}^{\numberofagents} z[b]_{i}^{j}$, $\sigma[b]_{i} = \sum_{j=1}^{s} p[b]_{j}/s$ and if there exists an integer $v$ such that $|\sigma[b]_{i} - p_{v}| < \frac{1}{2e^{2} \numberofagents}$, 
        where: $$p_{v} - \frac{1}{2} \left( \frac{\numberofagents-2}{\numberofagents} \right)^{v} \left( \left( \frac{\numberofagents}{\numberofagents-2} \right)^{v} - 1 \right),$$ 
        then $y[b] = v$.

If, for any $b$, no value $v$ exists or $y[0]+y[1] \neq \numberofagents$, the protocol aborts.

    \label{prot:vote}
\end{protocol}
\end{protocolfloat}

\end{document}

%% file: packages.tex
\usepackage{graphicx,graphics,times,bm,bbm,bbold,amssymb,amsmath,amsfonts, amsthm, dsfont,hyperref,color}
\usepackage[left=2cm,right=2cm,top=2cm,bottom=2cm]{geometry}
\usepackage[dvipsnames]{xcolor}
\usepackage[english]{babel}
\usepackage[autostyle]{csquotes}
\usepackage{relsize}
\usepackage[mode=buildnew]{standalone}
\usepackage{cleveref}
\usepackage{physics}
\usepackage[colorinlistoftodos]{todonotes}
\usepackage{soul}
\usepackage[many]{tcolorbox}
\usepackage{tkz-berge}
\usepackage{wrapfig}
\usepackage{nicefrac}

%% file: commands.tex
\definecolor{main}{HTML}{5989cf}    
\definecolor{sub}{HTML}{cde4ff}     

\tcbset{
    sharp corners,
    colback = white,
    before skip = 0.2cm,    
    after skip = 0.5cm      
}                           

\newtcolorbox{blue_box}[1]{
    title = #1,
    fonttitle = \bf,
    colback = sub, 
    colframe = main, 
    boxrule = 0pt, 
    toprule = 3pt, 
    bottomrule = 3pt 
}

\newtcolorbox{protocol_box}[1]{
    title = #1,
    fonttitle = \bf,
    colframe = gray!75,
    colback = white, 
    boxrule = 1pt, 
    toprule = 3pt, 
    bottomrule = 3pt, 
    breakable
}

\newcommand{\alice}{\ensuremath{\mathcal{A}}}

\NewDocumentCommand\GHZstate{O{n}}{\ensuremath{\ket{\mathrm{GHZ}_{#1}}}}
\NewDocumentCommand\GHZstatebra{O{n}}{\ensuremath{\bra{\mathrm{GHZ}_{#1}}}}
\NewDocumentCommand\GHZstateketbra{O{n}}{\ensuremath{\ketbra{\mathrm{GHZ}_{#1}}{\mathrm{GHZ}_{#1}}}}
\newcommand{\GHZtext}{\ensuremath{\mathrm{GHZ}}}

\NewDocumentCommand\GHZstatetheta{O{n}O{\avgp}}{\ensuremath{\ket{\mathrm{GHZ}_{#1}(#2)}}}
\NewDocumentCommand\GHZstatethetabra{O{n}O{\avgp}}{\ensuremath{\bra{\mathrm{GHZ}_{#1}(#2)}}}
\NewDocumentCommand\GHZstatethetaketbra{O{n}O{\avgp}}{\ensuremath{\ketbra{\mathrm{GHZ}_{#1}(#2)}{\mathrm{GHZ}_{#1}(#2)}}}

\newcommand{\avgp}{\ensuremath{\bar\theta_\p}}
\newcommand{\dis}{\mathcal{D}}
\newcommand{\hon}{\ensuremath{\mathcal{D}^{c}}}
\newcommand{\g}{\mathcal{G}}
\newcommand{\p}{\ensuremath{\mathcal{P}}}
\newcommand{\tar}{\ensuremath{\mathcal{T}}} 
\newcommand{\np}{\ensuremath{\overline{\mathcal{P}}}}
\newcommand{\e}{\mathrm{e}}
\newcommand{\numberofagents}{n}
\newcommand{\sizeofp}{m}
\newcommand{\sizeofd}{m_{d}}

\newcommand{\distributedghzstate}{N}
\newcommand{\numberoftotalrounds}{L}

\newcommand{\si}[1][]{\sigma^{\mathrm{in}}_{#1}}
\newcommand{\sout}[1][]{\sigma^{\mathrm{out}}_{#1}}
\newcommand{\soutdis}[1][]{\sigma^{\mathrm{out},\dis}_{#1}}
\newcommand{\ri}[1][]{\rho^{\mathrm{in}}_{#1}}
\newcommand{\ro}[1][]{\rho^{\mathrm{out}}_{#1}}
\newcommand{\rodis}[1][]{\rho^{\mathrm{out},\dis}_{#1}}

\newcommand{\prch}{\Gamma}

\newcommand{\reg}{{PTKCE}}
\newcommand{\regg}{{P_\g T_\g K_\g C_\g E_\g}}

\newcommand{\vecp}{\Vec{\theta}}
\newcommand{\QFIm}{\mathcal{Q}}
\newcommand{\quantreg}{\ensuremath{R}}

\newcommand{\yes}{\centering\textcolor{blue}{\checkmark}}
\newcommand{\no}{\centering\textcolor{red}{$\times$}}

\newcommand{\NOTIFICATION}{\texttt{NOTIFICATION}}

\newcommand{\PARITY}{\texttt{PARITY}}
\newcommand{\VOTE}{\texttt{VOTE}}

\newcommand{\STAVER}{\texttt{STATE VERIFICATION}}
\newcommand{\STAVERshort}{\texttt{SV}}

\newcommand{\PARVER}{\texttt{PARITY VERIFICATION}}
\newcommand{\PARVERshort}{\texttt{PV}}

\newcommand{\PAREST}{\texttt{PARITY ESTIMATION}}
\newcommand{\PARESTshort}{\texttt{PE}}

\newcommand{\ACKA}{\texttt{ANONYMOUS CONFERENCE KEY AGREEMENT}}
\newcommand{\ACKAshort}{\texttt{ACKA}}

\newcommand{\APPE}{\texttt{ANONYMOUS PRIVATE PARAMETER ESTIMATION}}
\newcommand{\APPEshort}{\texttt{APPE}}

\newtheorem{definition}{Definition}
\newtheorem{lemma}{Lemma}
\newtheorem*{remark}{Remark}

\renewcommand{\leq}{\leqslant}
\renewcommand{\geq}{\geqslant}

%% file: protocolenvironment.tex
\usepackage{enumitem}
\usepackage{booktabs}
\usepackage{newfloat}
\usepackage{etoolbox}
\usepackage{tcolorbox}

\DeclareFloatingEnvironment[fileext=frm,placement={!ht},name=Frame]{protocolfloat}

\newcounter{protocol}
\renewcommand{\theprotocol}{\arabic{protocol}}

\newenvironment{protocol}[3]{
\par\addvspace{\topsep}\refstepcounter{protocol}
\begin{tcolorbox}[
            halign title=left,
            colback=black!5!white,
            colframe=black!75!,
            rounded corners,
            title=\textbf{Protocol \theprotocol \;-\;} #1,
            parbox=false
            ]
\noindent
    \begin{tabular}{rp{0.7\linewidth}}
      \ifstrempty{#2}{}{\texttt{Input:} & #2 \\}
      \texttt{Goal:} & #3 \\
  \end{tabular} \\
  \tcblower
      \raggedright
    
}{
\vspace{-5pt}
\end{tcolorbox}
\vspace{-10pt}
}

\crefname{protocol}{Protocol}{Protocols}
\Crefname{protocol}{Protocol}{Protocols}

\newlist{protocollist}{enumerate}{1}
\setlist[protocollist,1]{label*=\arabic*:, ref=\arabic*}
\crefname{protocollist}{step}{steps}
\crefalias{protocollisti}{protocollist}
\Crefname{protocollist}{Step}{Steps}